\documentclass[a4paper,11pt]{article}

\usepackage{extarrows}
\usepackage[numbers,sort&compress]{natbib}
\usepackage[all]{xy}
\usepackage{amsmath,amsfonts,amssymb,amsthm,epsfig,amscd,comment,latexsym,psfrag}
\usepackage{CJK}
\usepackage{mathrsfs}
\usepackage{nicefrac,xspace,tikz}
\usepackage{arydshln}
\usetikzlibrary{arrows}
\usepackage{graphicx}
\usepackage{caption,subcaption}
\usepackage{xcolor,float}
\usepackage{appendix}
\usepackage{graphicx}
\usepackage{caption}
\makeatletter
\newcommand{\rmnum}[1]{\romannumeral #1}
\newcommand{\Rmnum}[1]{\expandafter\@slowromancap\romannumeral #1@}
\makeatother
\def\footnoterule{\kern 1mm \hrule width 7cm \kern 2.2mm}%

\def\dsum{\displaystyle\sum}

\def\Tr{\mathrm{Tr}}
\def\tr{\mathrm{tr}}

\renewcommand{\b}{\beta}

\usepackage{graphicx,amssymb,amsmath,bm,latexsym}
\allowdisplaybreaks

\begin{document}
\begin{titlepage}
\begin{center}
{\Large\bf $W$-representations of the fermionic matrix and Aristotelian tensor models}\vskip .2in
{\large Lu-Yao Wang$^{a}$, Rui Wang$^{b}$, Ke Wu$^{a}$,
Wei-Zhong Zhao$^{a,}$\footnote{Corresponding author: zhaowz@cnu.edu.cn}} \vskip .2in
$^a${\em School of Mathematical Sciences, Capital Normal University,
Beijing 100048, China} \\
$^b${\em Department of Mathematics, China University of Mining and Technology,
Beijing 100083, China}\\

\begin{abstract}
We show that the fermionic matrix model can be realized by $W$-representation.
We construct the Virasoro constraints with higher algebraic structures, where
the constraint operators obey the Witt algebra and null 3-algebra.
The remarkable feature is that the character expansion of the partition function can be easily
derived from such Virasoro constraints.
It is a $\tau$-function of the KP hierarchy.
We construct the fermionic Aristotelian tensor model
and give its $W$-representation. Moreover, we analyze
the fermionic red tensor model and present the
$W$-representation and character expansion of
the partition function.

\end{abstract}

\end{center}

{\small Keywords: Matrix Models, Conformal and $W$ Symmetry}

\end{titlepage}

\section{Introduction}

Matrix models provide a rich set of approaches to physical systems.
Much interest has been attributed to their remarkable properties.
As is well known, the partition function of some matrix models,
such as the Gaussian Hermitian and complex matrix models \cite{Shakirov2009}-\cite{MironovJHEP082018},
Kontsevich matrix model \cite{Alexandrov2011}, can be presented as the forms of $W$-representations
and character expansions of Schur polynomials.
There is the superintegrability for these matrix models,
which means that the average of a properly chosen symmetric function is proportional to ratios of
symmetric functions on a proper locus, i.e., $<character>\sim character$.
For the cases of $\beta$- and $q,t$-deformed matrix models,
the character expansions of the partition functions are given
by the Jack and Macdonald polynomials, respectively \cite{Morozov2018}-\cite{Cassia2020}.
The $W$-representation gives a dual expression for partition function
through differentiation rather than integration.
The $W$-representations in terms of characters were analyzed in \cite{Alexandrov2014,MironovJHEP082018}.
The constraints for matrix models enable us to effectively analyze the structures of matrix models.
There have already been many works along this direction. Recently the remarkable works are devoted to analyze
the character expansion of the Gaussian Hermitian model
by using the Virasoro constraints \cite{AMironov2104,AMironov2105}.

Matrix models are associated with discretized random surfaces and $2D$ quantum
gravity. Tensor models are the generalizations of matrix models from matrices to tensor,
which are originally introduced to describe the higher dimensional quantum gravity
\cite{Jonsson}-\cite{Sasakura}.
A few methods which allow one to connect calculations in the tensor models to those
in the matrix models were introduced in \cite{ItoyamaJHEP2017}-\cite{Itoyama1703}.
The $W$-representations and character expansions of the partition functions
have been extended to (rainbow) tensor models \cite{ItoyamaPLB2019}-\cite{Kang2021}.

The fermionic matrix models involve matrices with anticommuting elements \cite{Makeenko1994}-\cite{GordonW.Semenof}.
A well known fermionic one-matrix model is given by \cite{GordonW.Semenof}
\begin{eqnarray}\label{partition1}
Z=\int d\psi d\bar{\psi} \exp[N^2\tr(\bar{\psi} \psi +\sum_{k\geq 0}g_{k}(\bar{\psi} \psi)^{k})],
\end{eqnarray}
where~$\psi$ and~$\bar{\psi}$ are independent complex Grassmann-valued~$N\times N$ matrices,
the integration measure in (\ref{partition1}) is the Haar measure on the Grassmann algebra,~
$d\psi d\bar \psi=\prod_{i,j}d\psi_{i,j} d\bar \psi_{i,j}$, which is defined by the Berezin rules for integrating
Grassmann variables. The fermionic one-matrix model (\ref{partition1}) can be viewed as a $D=0$
dimensional quantum field theory of a self-interacting Dirac fermion.
Note that the matrices with anticommuting elements cannot be diagonalized by unitary transformations.
Unlike the Hermitian matrix models, the fermionic matrix model can not be expressed as the eigenvalue model.
However, it can be analyzed by methods similar to those used for Hermitian
matrix models.  It was showed that the loop equations of the adjoint fermion matrix model
are identical to those for the Hermitian one-matrix model with generalized Penner potential.

The fermionic matrix models are also generalized to the fermionic tensor cases \cite{Robert,Shiroman}.
It seems that the fermionic tensor model rather than the bosonic tensor model
is more relevant to the problem of understanding holography.
The gauge invariants of fermionic tensor model can be constructed
by means of the representation theory. There are the different counting formulas
for the number of gauge invariant operators in bosonic and fermionic tensor models \cite{Robert}.

Already a considerable amount is known about the $W$-representations of matrix and tensor models.
Recently the corresponding analysis has been carried out for the supereigenvalue models  \cite{Chen,wang2020}.
The aim of the present paper is to make an attempt of studying
$W$-representations of fermionic matrix and tensor models.
More precisely, we focus on the fermionic matrix and Aristotelian tensor models and present
their $W$-representations.

This paper is organized as follows. In section 2, we analyze the fermionic matrix model and give its $W$-representation.
Furthermore, we derive the character expansion of the partition function and compact expressions of the
correlators from the Virasoro constraints. In section 3, we construct the fermionic Aristotelian tensor model
and give its $W$-representation. In section 4, we give $W$-representation of the fermionic red tensor model.
Moreover we show that there is the character expansion for the partition function.
We end this paper with the conclusions in section 5.

\section{$W$-representation and character expansion of the fermionic matrix model}
\subsection{$W$-representation of the fermionic matrix model}
Let us consider the fermionic matrix model
\begin{eqnarray}\label{FMM}
Z_F=\frac{\int d\psi d\bar{\psi} \exp[N\sum_{k>0}\frac{p_k}{k}\tr(\bar{\psi} \psi)^k
+N^2\tr(\bar{\psi} \psi)]}{\int d\psi d\bar\psi\exp(N^2\tr\bar\psi\psi)},
\end{eqnarray}
where~all traces are normalized as~$\tr A=\frac{1}{N}\sum_{i=1}^N A_{ii}$.

The following Virasoro constraints can be derived from (\ref{FMM}) by implementing invariance
under the infinitesimal shifts
$\psi\rightarrow \psi+\epsilon\psi(\bar\psi \psi)^n\ (n\geq0)$, $\bar\psi\rightarrow \bar\psi$,
\begin{eqnarray}\label{vira}
\mathcal{L}_n Z_F=0,
\end{eqnarray}
where
\begin{eqnarray}\label{viraconst}
\mathcal{L}_n =k(n-k)\sum_{k=1}^{n-1}\frac{\partial^2}{\partial p_{k}\partial p_{n-k}}-\delta_{n,0}N^2
+\sum_{k>0}N(n+k)p_k\frac{\partial}{\partial p_{n+k}}
+(n+1)N\frac{\partial}{\partial p_{n+1}}.
\end{eqnarray}
The constraint operators (\ref{viraconst}) obey the Witt algebra
\begin{eqnarray}\label{witt}
[\mathcal{L}_m,\mathcal{L}_n]=(m-n)\mathcal{L}_{m+n}.
\end{eqnarray}
However, straightforward calculation shows that they do not yield the closed higher algebraic structures.

Let us now take the change of variables given by
\begin{eqnarray}\label{change}
\psi\longrightarrow \psi+\epsilon\sum_{n=0}^{\infty}\frac{p_{n+1}}{N}\psi(\bar\psi \psi)^n,\ \bar\psi\longrightarrow \bar\psi.
\end{eqnarray}
By requiring that the partition function is invariant under the infinitesimal transformations (\ref{change}),
it leads to the constraint
\begin{eqnarray}
(\hat{D}-\hat{W})Z_F=0,
\end{eqnarray}
where the operators $\hat{D}$ and $\hat{W}$ are given by
\begin{eqnarray}\label{D&W}
&&\hat{D}=\sum_{n=1}^{\infty}np_{n}\frac{\partial}{\partial p_{n}},
\nonumber\\
&&\hat{W}=Np_1-\frac{1}{N}\sum_{n,k=1}^{\infty}(n+k-1)p_np_k
\frac{\partial}{\partial p_{n+k-1}}
-\frac{1}{N}\sum_{n,k=1}^{\infty}nkp_{n+k+1}\frac{\partial^2}{\partial p_{k}\partial p_{n}}.
\end{eqnarray}

To explore the properties of $\hat D$ and $\hat W$, we rewrite (\ref{FMM}) as
\begin{eqnarray}\label{partition1-deg}
Z_F=\sum_{s=0}^{\infty}Z_F^{(s)},
\end{eqnarray}
where
\begin{eqnarray}\label{partition1-deg2}
Z_F^{(s)}=\sum_{l=0}^{\infty}\sum_{k_1+\cdots k_l=s}^{\infty}
\frac{N^{l}}{l!} \frac{p_{k_1}\cdots p_{k_l}}{k_1\cdots k_l}\langle \prod_{n=1}^l\tr(\bar\psi \psi)^{k_n} \rangle,
\end{eqnarray}
and $\langle \prod_{n=1}^l\tr(\bar\psi \psi)^{k_n} \rangle$ are the correlators defined by
\begin{eqnarray}\label{defcorrelator}
\langle \prod_{n=1}^l\tr(\bar\psi \psi)^{k_n} \rangle=\frac{\int d\psi d\bar \psi \prod_{n=1}^l\tr(\bar\psi \psi)^{k_n} \exp(N^2 \tr \bar\psi \psi)}
{\int d\psi d\bar \psi  \exp(N^2 \tr \bar\psi \psi)}.
\end{eqnarray}

Through the operators $\hat D$ and $\hat W$ acting on $Z^{(s)}$
\begin{eqnarray}\label{dzs}
\hat {D} Z_F^{(s)}=sZ_F^{(s)},
\end{eqnarray}
\begin{eqnarray}\label{increa}
\hat{W}Z_F^{(s)}=(s+1)Z_F^{(s+1)},
\end{eqnarray}
we see that $\hat D$ and $\hat{W}$
are the operators preserving and increasing the grading, respectively.
The commutator of $\hat D$ with $\hat W$ is
\begin{eqnarray}
[\hat D,\hat W]=\hat W.
\end{eqnarray}
In terms of the operator $\hat{W}$, the partition function (\ref{FMM})
can be realized by the $W$-representation
\begin{eqnarray}\label{wrepresentation}
Z_F=\exp(\hat{W})\cdot 1,
\end{eqnarray}
which is the dual expression of (\ref{FMM}) through differentiation rather than integration.

Let us write the higher power of $\hat{W}$ as
\begin{eqnarray}\label{W^m}
\hat{W}^m=\sum_{\substack{i_1+\cdots+i_l\\-j_1-\cdots-j_k=m}}
\mathcal{P}_{j_1,\cdots ,j_k}^{i_1,\cdots ,i_l}p_{i_1}\cdots p_{i_l}\frac{\partial}{\partial p_{j_1}}\cdots\frac{\partial}{\partial p_{j_k}},
\end{eqnarray}
where ~$\mathcal{P}_{j_1,\cdots ,j_k}^{i_1,\cdots ,i_l}$ are the expansion coefficients.

Then it is not difficult to give the correlators
from the $W$-representation (\ref{wrepresentation})
\begin{eqnarray}\label{correlator}
\langle \prod_{n=1}^l\tr(\bar\psi \psi)^{i_n} \rangle
=\frac{l!}{m!\lambda_{(i_1,\cdots,i_l)}}
\sum_{k=0}^m\sum_{\tau}\mathcal{P}^{\tau(i_1),\cdots,\tau(i_l)}_{\underbrace{0\cdots0}_k}N^{k-l},
\end{eqnarray}
where~$m=i_1+\cdots+i_l$, $\tau$ denotes all distinct permutations of
$(i_1,\cdots,i_l)$ and~$\lambda_{(i_1,\cdots,i_l)}$ is the number of $\tau$ with respect to~$(i_1,\cdots,i_l)$.

Note that under the chiral transformation~$\psi\rightarrow\bar\psi,\ \bar\psi\rightarrow-\psi$,
the traces transform
as~$\tr(\bar\psi\psi)^k\rightarrow(-1)^{k+1}\tr(\bar\psi\psi)^k$.
It is easy to know that all the even correlators vanish
\begin{eqnarray}
\langle\tr(\bar\psi\psi)^{2k}\rangle=0.
\end{eqnarray}
Here the chiral transformation is the analog of the reflection
of the hermitian matrix $\phi\rightarrow-\phi$~in the Gaussian Hermitian matrix model,
which makes all the odd moments vanishing,~i.e., $\langle\tr(\bar\phi\phi)^{2k+1}\rangle=0$.

Let us give several exact correlators from (\ref{correlator})
\begin{eqnarray}
&&\langle\tr(\bar\psi\psi)\rangle=1,\nonumber\\
&&\langle\tr(\bar\psi\psi)\tr(\bar\psi\psi)\rangle=1-N^{-2},\nonumber\\
&&\langle\tr(\bar\psi\psi)\tr(\bar\psi\psi)\tr(\bar\psi\psi)\rangle
=1-2N^{-2}+2N^{-4},\nonumber\\
&&\langle\tr(\bar\psi\psi)^3\rangle=1-N^{-2},\nonumber\\
&&\langle\tr(\bar\psi\psi)\tr(\bar\psi\psi)^2\rangle=0.
\end{eqnarray}

\subsection{Character expansion from the Virasoro constraints}

By means of the operators $\hat D$ and $\hat W$,
we may construct the Virasoro constraints
\begin{eqnarray}\label{Vconstraint1}
L_m Z_F=0,
\end{eqnarray}
where the constraint operators are given by
\begin{eqnarray}\label{Virasoro constraint}
L_m=\hat{W}^m(\hat{W}-\hat{D}),
\end{eqnarray}
which yield the Witt algebra (\ref{witt}) and null 3-algebra
\begin{eqnarray}\label{3Valg}
[L_k, L_m, L_n]=0.
\end{eqnarray}

There is a significant difference between the Virasoro constraint operators
(\ref{viraconst}) and (\ref{Virasoro constraint}).
From the Virasoro constraints (\ref{Vconstraint1}), we can derive the correlators
(\ref{correlator}) as well.
There are the recursive relations of correlators from the Virasoro constraints (\ref{vira}).
We can calculate the correlators from these recursive relations. However, it is hard to give
the compact expression (\ref{correlator}).

To achieve the character expansion of the partition function,
let us rewrite (\ref{FMM}) as
\begin{eqnarray}\label{FMM-1}
Z_F=\sum_{R}\langle\chi_R\{\tr(\bar\psi\psi)^k\}\rangle\chi_R\{p\}=\sum_{R}C_R\chi_R,
\end{eqnarray}
where~$R=\{R_1,\cdots,R_{l_{(R)}}\}$~are the Young diagrams of the given size (number of boxes)~$|R|=\sum_iR_i$~
and length~$ l_{(R)}$,~$\chi_R$~is the Schur polynomial, and we have used the Cauchy formula
$\exp({\sum_{k=1}^{\infty}\frac{1}{k}a_kb_k})=\sum_{R}\chi_R\{a\}\chi_R\{b\}$.

The Virasoro constraints (\ref{Vconstraint1}) then become
\begin{eqnarray}\label{Vconstraint2}
\sum_R C_R \hat{W}^{m+1}\cdot\chi_R=\sum_{R}\sum_{|R|=m+1}^{\infty}|R|(|R|-1)\cdots(|R|-m)C_R\chi_R.
\end{eqnarray}
In order to show that there is the character expansion for the partition function,
the key point depends on the action result of $\hat{W}^{m+1}\cdot\chi_R$.

Since the operator $\hat{W}$ acting on $\chi_R$ gives
\begin{eqnarray}
\hat{W}\cdot\chi_R=-\frac{1}{N}\sum_{R+\square}[(j_\square-i_\square)^2-N^2]\chi_{R+\square},
\end{eqnarray}
here $(i_\square, j_\square)$~are the coordinates of the squares added to the Young diagram $R$,
we have
\begin{eqnarray}\label{wchiR}
\hat{W}^{m+1}\cdot\chi_R
%&=&\frac{{(-1)}^{m+1}}{N^{m+1}}\sum_{R+\square_1+\cdots+\square _{m+1}}
%[(j_{\square_1}-i_{\square_1})^2-N^2][(j_{\square_2}-i_{\square_2})^2-N^2]
%\nonumber\\&& \cdots[(j_{\square_{m+1}}-i_{\square_{m+1}})^2-N^2]\chi_{R+\square_1+\cdots+\square_{m+1}}\nonumber\\
=\frac{{(-1)}^{m+1}}{N^{m+1}}\sum_{S}n(S)\prod_{(i_\square, j_\square)\in S}[(j_{\square}-i_{\square})^2-N^2]\chi_{R+S},
\end{eqnarray}
where~$S$ is skew Young diagram with $|S|=m+1$,~$n(S)$ is the number of standard Young tableau with shape~$S$.

Due to (\ref{wchiR}), there are the recursive relations from (\ref{Vconstraint2})
\begin{eqnarray}\label{recursive}
C_{S+R}=
\frac{{(-1)}^{m+1}n(R)}{N^{m+1}}\frac{\prod_{(i_\square, j_\square)\in R}[(j_{\square}-i_{\square})^2-N^2]}
{(|S|+m+1)(|S|+m)\cdots(|S|+1)}C_S,
\end{eqnarray}
where $|R|=m+1$, we have interchanged the indices $R$ and $S$ for later convenience.

The explicit form of $C_{R}$ is obtained from (\ref{recursive})
with initial value $C_{\varnothing}=1$
\begin{eqnarray}\label{CS-2}
C_{R}=\frac{{(-1)}^{m+1}n(R)}{N^{m+1}}\frac{\prod_{(i_\square, j_\square)\in R}[(j_{\square}-i_{\square})^2-N^2]}
{(m+1)!}.
\end{eqnarray}

The coefficient $n(R)$ in (\ref{CS-2}) also represents the dimension of the irreducible representation~
$\pi_R$ of symmetry group~$S_{|R|}$,
i.e.,~$n(R)=\frac{(m+1)!}{\prod_{x\in R}hook(x)}$, where $hook(x)$ is the number of boxes in hook.

By using the hook formulas \cite{macdonal,Fulton}
\begin{eqnarray}\label{hook1}
\frac{D_R(N)}{d_R}=\prod_{(i_\square, j_\square)\in R}(N-i_{\square}+j_{\square})
,
\end{eqnarray}
and
\begin{eqnarray}\label{hook2}
d_R
%&=&\frac{\prod_{i,j}(R_i-R_j-i+j)}{\prod_{i=1}^{m}(R_i+m-i)!}\nonumber\\&=&\frac{\dim \pi_R}{(m+1)!}
=\frac{1}{\prod_{x\in R}hook(x)},
\end{eqnarray}
where~$D_R(N)=\chi_R\{p_k=N\}$ and $d_R=\chi_R\{p_k=\delta_{k,1}\}$~are respectively the dimension of representation~$R$~
for the linear group~$GL(N)$ and symmetric group $S_{|R|}$~divided by~$|R|!$, we obtain the final expression of (\ref{CS-2})
\begin{eqnarray}\label{C_S}
C_R=(\frac{-1}{N})^{|R|}\frac{D_R(N)D_R(-N)}{d_R}.
\end{eqnarray}

Thus we reach the desired result
\begin{eqnarray}\label{character1}
Z_F=\exp(\hat{W})\cdot 1
%=\sum_{n=0}^{\infty}\frac{1}{n!}W^n\cdot1
=\sum_RC_R\chi_R.
\end{eqnarray}

For any partition function of the form \cite{AndreiMironov}
\begin{eqnarray}\label{tau}
Z_\omega=\sum_R \chi_R\{p\}\chi_R\{\bar p\}\omega_R
\end{eqnarray}
with the function $\omega_R=\Pi_{(i,j)\in R}f(i-j)$
and ${\bar p}_k$ just arbitrary parameters,
it is a $\tau$-function of the KP hierarchy \cite{Orlov2001}-\cite{Kharchev1995}.
Since (\ref{character1}) coincides with (\ref{tau}), we conclude that
the character expansion of the fermionic matrix model is indeed a $\tau$-function of the KP hierarchy.

We note that the action of $\hat{W}^n\cdot1$ can be written as
\begin{eqnarray}\label{hatwn1}
\hat{W}^n\cdot1
&=&\sum_{\lambda\mapsto n}\sum_{\sigma\in S_l}\mathcal{P}^{\sigma{(\lambda_1)},\cdots ,\sigma{(\lambda_l)}}p_{\lambda_1}\cdots p_{\lambda_l},
\nonumber\\
&=&\sum_{\lambda,R\mapsto n}\sum_{\sigma\in S_l}\mathcal{P}^{\sigma{(\lambda_1)},\cdots ,\sigma{(\lambda_l)}}\psi_{R}(\lambda)\chi_{R}\{p\},
\end{eqnarray}
where $\lambda=(\lambda_1,\cdots,\lambda_l)$ is a partition of $n$,~$\sigma$~is a permutation
in the symmetry group~$S_l$~,~$\psi_{R}(\lambda)$~is the symmetric group character.
In deriving the above action, we have used the formula
$p_\lambda=p_{\lambda_1}\cdots p_{\lambda_l}=\sum_{R\mapsto n}\psi_{R}(\lambda)\chi_{R}\{p\}$.

By means of (\ref{hatwn1}), it is easy to give another expression for $C_R$ from (\ref{character1})
\begin{eqnarray}\label{C_R}
C_R=\frac{1}{|R|!}\sum_{\substack{ \lambda=(\lambda_1,\cdots,\lambda_l)\\ \lambda\mapsto |R|}}
\sum_{ \sigma\in S_l}\mathcal{P}^{\sigma{(\lambda_1)},\cdots ,\sigma{(\lambda_l)}}\psi_{R}(\lambda).
\end{eqnarray}

\section{$W$-representation of the fermionic Aristotelian tensor model}

The Aristotelian tensor model with a single complex tensor of rank 3 and the RGB (red-green-blue)
symmetry is the simplest of the rainbow tensor models \cite{ItoyamaJHEP2017}.
It can be realized by $W$-representation. In the previous section, we have analyzed
the fermionic matrix model. Let us now turn to the fermionic generalization of Aristotelian tensor model.

We introduce the gauge-invariant operators of level~$n$
\begin{eqnarray}\label{ksigma}
\mathcal{K}_{\sigma}^{(n)}=
\mathcal{K}_{\sigma_1\otimes \sigma_2\otimes\sigma_3}^{(n)}= \prod_{p=1}^{n}
\bar\Psi_{\textcolor{red}{i}^{(p)}}^{\textcolor{green}{j_1}^{(p)},\textcolor{blue}{j_2}^{(p)}}
 \Psi_{\textcolor{green}{j_1}^{\sigma_2(p)},\textcolor{blue}{j_2}^{\sigma_3(p)}}^{\textcolor{red}{i}^{\sigma_1(p)}},
\end{eqnarray}
where the gauge symmetry is $U(\textcolor{red}{N_1})\otimes U(\textcolor{green}{N_2})\otimes U(\textcolor{blue}{N_3})$,
$\bar{\Psi}^{\textcolor{red}{i}}_{\textcolor{green}{j_1},\textcolor{blue}{j_{2}}}$
and $\Psi_{\textcolor{red}{i}}^{\textcolor{green}{j_1},\textcolor{blue}{j_{2}}}$ are the
fermionic tensors of rank~$3$ with one covariant and two contravariant indices  which are assigned with different color, $\sigma$
is an element of the double coset $\mathcal{S}_n^3= S_n\backslash S_n^{\otimes 3}/S_n$.
The fermions are described by Grassmann variables. There is a natural action of~$\mathcal{S}_n^3$ on~
$U(\textcolor{red}{N_1})\otimes U(\textcolor{green}{N_2})\otimes U(\textcolor{blue}{N_3})$ defined as follows:
\begin{eqnarray}
%\sigma^{'}
\xi\cdot
(\bar\Psi_{\textcolor{red}{i}}^{\textcolor{green}{j_1},\textcolor{blue}{j_2}}
 \Psi_{\sigma_2(\textcolor{green}{j_1}),\sigma_3(\textcolor{blue}{j_2})}^{\sigma_1(\textcolor{red}{i})})
=\bar\Psi_{\xi(\textcolor{red}{i})}^{\xi(\textcolor{green}{j_1}),\xi(\textcolor{blue}{j_2})}
 \Psi_{\xi(\sigma_2(\textcolor{green}{j_1})),\xi(\sigma_3(\textcolor{blue}{j_2}))}^{\xi(\sigma_1(\textcolor{red}{i}))}
=sgn(\xi)\bar\Psi_{\textcolor{red}{i}}^{\textcolor{green}{j_1},\textcolor{blue}{j_2}}
 \Psi_{\sigma_2(\textcolor{green}{j_1}),\sigma_3(\textcolor{blue}{j_2})}^{\sigma_1(\textcolor{red}{i})},
\end{eqnarray}
where $\xi$  is a permutation in $\mathcal{S}_n^3$.

In similarity with the Aristotelian tensor model, we may introduce
the cut operation $\Delta_F$ and join operation $\{ \ \}_F$ on the gauge-invariant operators
\begin{eqnarray}\label{cut}
\Delta_F \mathcal{K}_{\alpha}^{(a)}&=&\sum_{\textcolor{red}{i}=1}^{\textcolor{red}{N_1}}
\sum_{\textcolor{green}{j_1}=1}^{\textcolor{green}{N_2}}
\sum_{\textcolor{blue}{j_{2}}=1}^{\textcolor{blue}{N_{3}}}
\dfrac{\partial^2 \mathcal{K}_{\alpha}^{(a)}}{\partial \bar\Psi_{\textcolor{red}{i}}^{\textcolor{green}{j_1},\textcolor{blue}{j_{2}}}
\partial \Psi^{\textcolor{red}{i}}_{\textcolor{green}{j_1},\textcolor{blue}{j_{2}}}}\nonumber\\
&=&\sum_{k=1}^3\sum_{\substack{\beta_1,\cdots,\beta_k\\ b_1+\cdots+b_k+1=a}}\Delta_{\alpha}^{\beta_1,\cdots,\beta_k}
\mathcal{K}_{\beta_1}^{(b_1)}\cdots\mathcal{K}_{\beta_k}^{(b_k)},\ a\geqslant 2,
\end{eqnarray}
and
\begin{eqnarray}\label{join}
\{\mathcal{K}_{\sigma}^{(n)},\mathcal{K}_{\alpha}^{(a)}\}_F
&=&
\sum_{\textcolor{red}{i}=1}^{\textcolor{red}{N_1}}
\sum_{\textcolor{green}{j_1}=1}^{\textcolor{green}{N_2}}
\sum_{\textcolor{blue}{j_{2}}=1}^{\textcolor{blue}{N_{3}}}
\dfrac{\partial \mathcal{K}_{\sigma}^{(n)}}{\partial \bar\Psi_{\textcolor{red}{i}}^{\textcolor{green}{j_1},\textcolor{blue}{j_{2}}}}
\dfrac{\partial \mathcal{K}_{\alpha}^{(a)}}{ \partial \Psi^{\textcolor{red}{i}}_{\textcolor{green}{j_1},\textcolor{blue}{j_{2}}}}\nonumber\\
&=&
\sum_{\deg\beta=n+a-1}
\gamma_{\sigma,\alpha}^{\beta}
\mathcal{K}_{\beta}^{(n+a-1)},
\end{eqnarray}
where $\Delta_{\alpha}^{\beta_1,\cdots,\beta_k}$  and~$\gamma_{\sigma,\alpha}^{\beta}$ are coefficients.

It is known that the keystone operators play an important role in the Aristotelian
tensor model \cite{ItoyamaJHEP2017,Itoyama1710}. They may generate a graded ring of gauge
invariant operators  with addition, multiplication, cut and join operations.
The ring contains the so-called tree and loop operators.
When the operator belongs to the sub-ring generated only by the
join operation, this operator is the tree operator, otherwise, it is the loop operator.

For the case of fermionic Aristotelian tensor model, the ring is generated by the following
keystone operators
\begin{eqnarray}
%\textcolor{red}{\mathcal{K}_2}=\mathcal{K}_{(12)\otimes id\otimes id}=
%\bar\Psi_{\textcolor{red}{i}^{(1)}}^{\textcolor{green}{j_1}^{(1)},\textcolor{blue}{j_2}^{(1)}}
%\Psi^{\textcolor{red}{i}^{(2)}}_{\textcolor{green}{j_1}^{(1)},\textcolor{blue}{j_2}^{(1)}}
%\bar\Psi_{\textcolor{red}{i}^{(2)}}^{\textcolor{green}{j_1}^{(2)},\textcolor{blue}{j_2}^{(2)}}
%\Psi^{\textcolor{red}{i}^{(1)}}_{\textcolor{green}{j_1}^{(2)},\textcolor{blue}{j_2}^{(2)}}
\includegraphics[height=2cm]{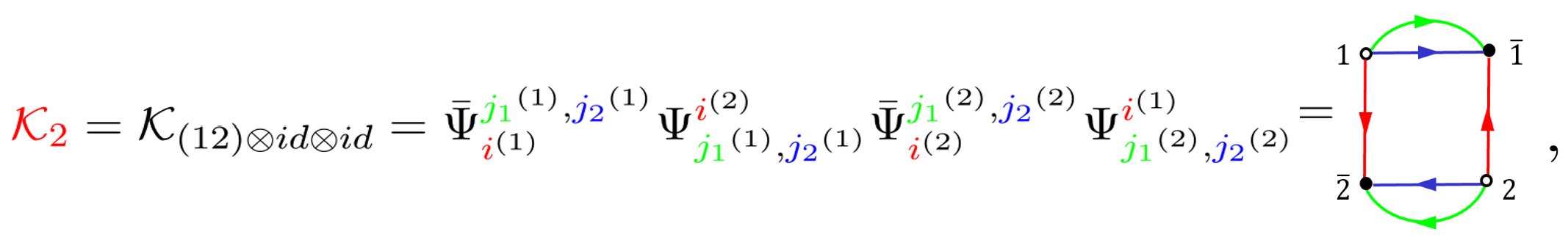}
\end{eqnarray}
\begin{eqnarray}
%\textcolor{green}{\mathcal{K}_2}=\mathcal{K}_{id\otimes(12) \otimes id}=
%\bar\Psi_{\textcolor{red}{i}^{(1)}}^{\textcolor{green}{j_1}^{(1)},\textcolor{blue}{j_2}^{(1)}}
%\Psi^{\textcolor{red}{i}^{(1)}}_{\textcolor{green}{j_1}^{(2)},\textcolor{blue}{j_2}^{(1)}}
%\bar\Psi_{\textcolor{red}{i}^{(2)}}^{\textcolor{green}{j_1}^{(2)},\textcolor{blue}{j_2}^{(2)}}
%\Psi^{\textcolor{red}{i}^{(2)}}_{\textcolor{green}{j_1}^{(1)},\textcolor{blue}{j_2}^{(2)}}
\includegraphics[height=1.9cm]{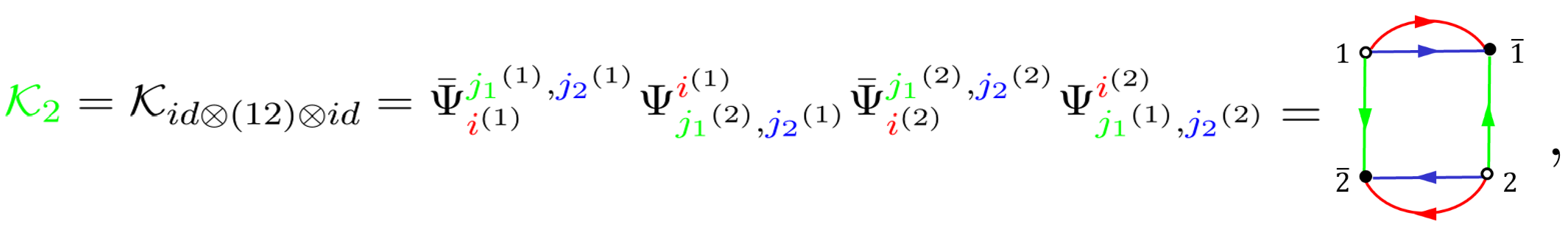}
\end{eqnarray}
where the number $1$, $\bar{1}$ and $2$, $\bar{2}$ represent the first two fermionic fields $\bar{\Psi}$ and $\Psi$,
and the last two fields $\bar{\Psi}$ and $\Psi$ respectively. The vertices are fields (tensors),
the different color thin lines represent the contraction of indices in the operators,
and the directions of arrows depend on the choice of covariant and contravariant indices.

For the tree operators and loop operators in the ring, they
have the same construction rules with the Aristotelian tensor model, which are chains of keystone operators connected
by Feynmann propagators. In this case, the Feynmann propagators are depicted as thick black lines with arrow,
where the arrow associates a factor of $\pm 1$ to the contracting direction of the two fields.

\begin{figure}[H]
\centering
\begin{minipage}[c]{0.4\textwidth}
\centering
\includegraphics[height=1.9cm]{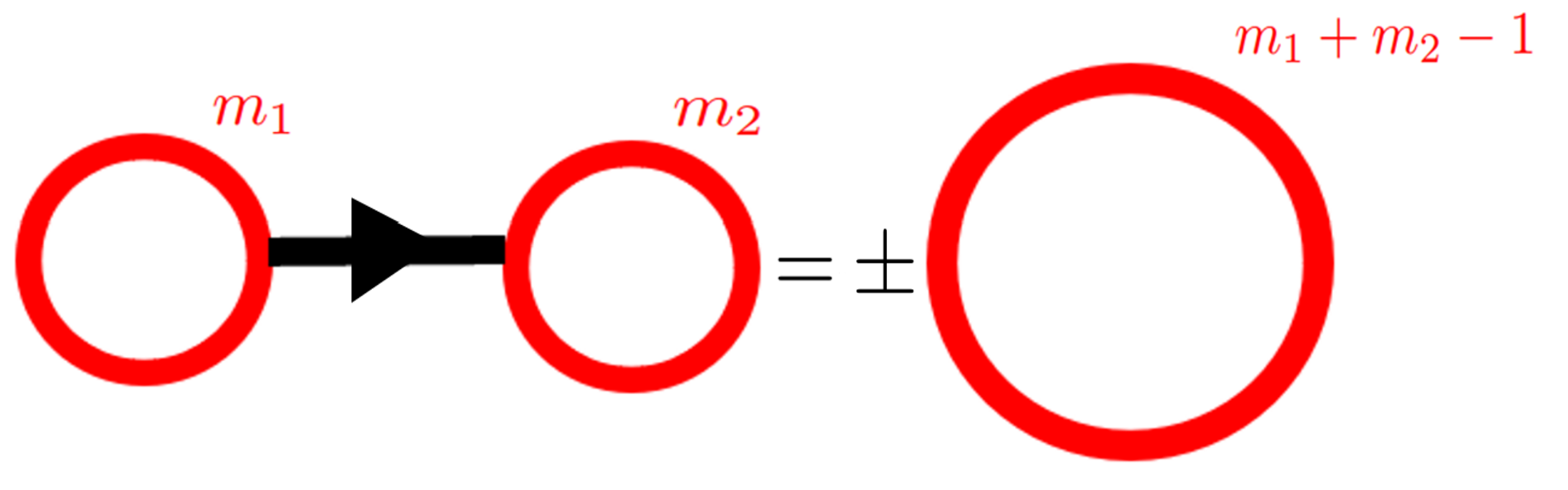}
\caption*{(a)}
\end{minipage}
\begin{minipage}[c]{0.4\textwidth}
\centering
\includegraphics[height=2cm]{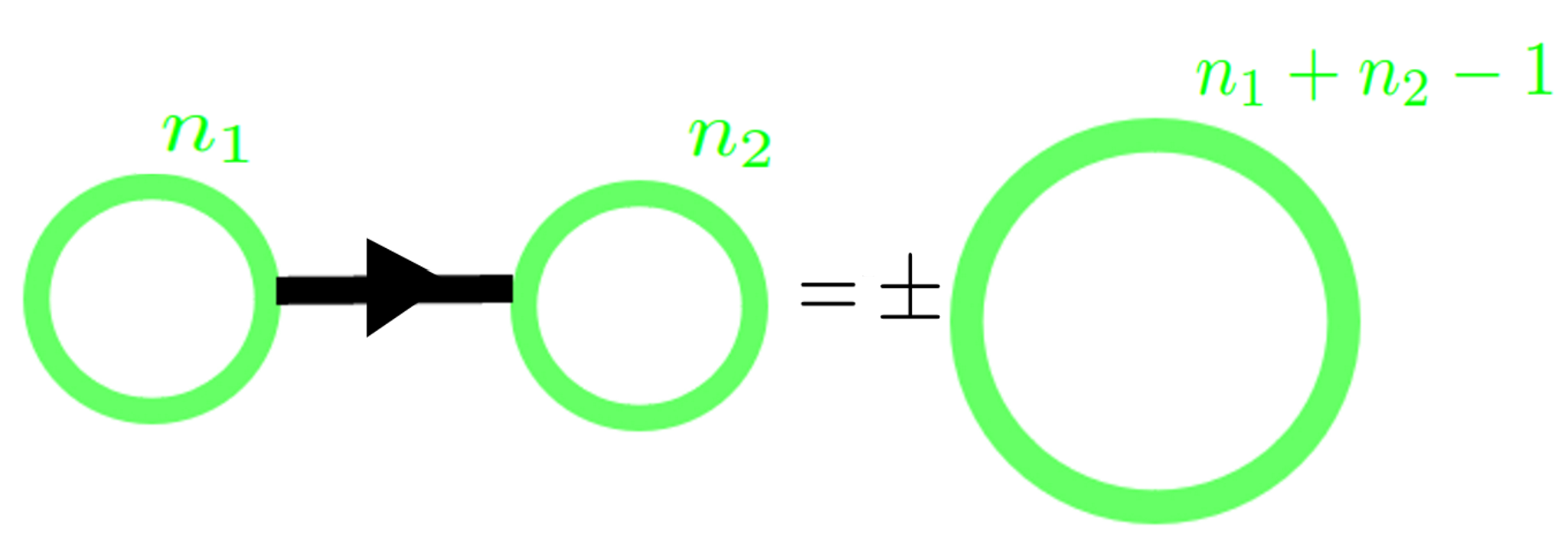}
\caption*{(b)}
\end{minipage}
\nonumber\\
\begin{minipage}[c]{0.4\textwidth}
\centering
\includegraphics[height=2.3cm]{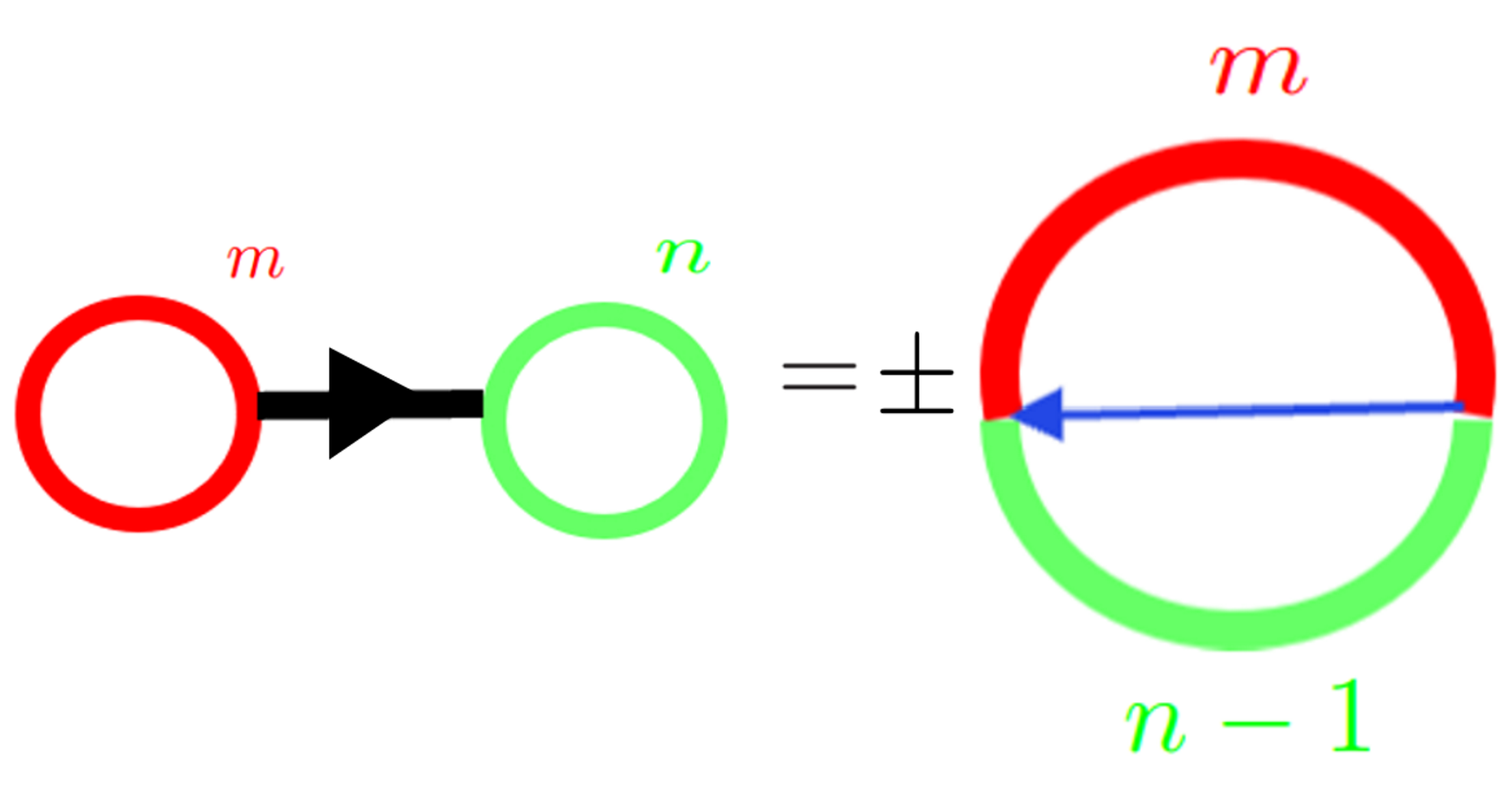}
\caption*{(c)}
\end{minipage}
\begin{minipage}[c]{0.4\textwidth}
\centering
\includegraphics[height=2cm]{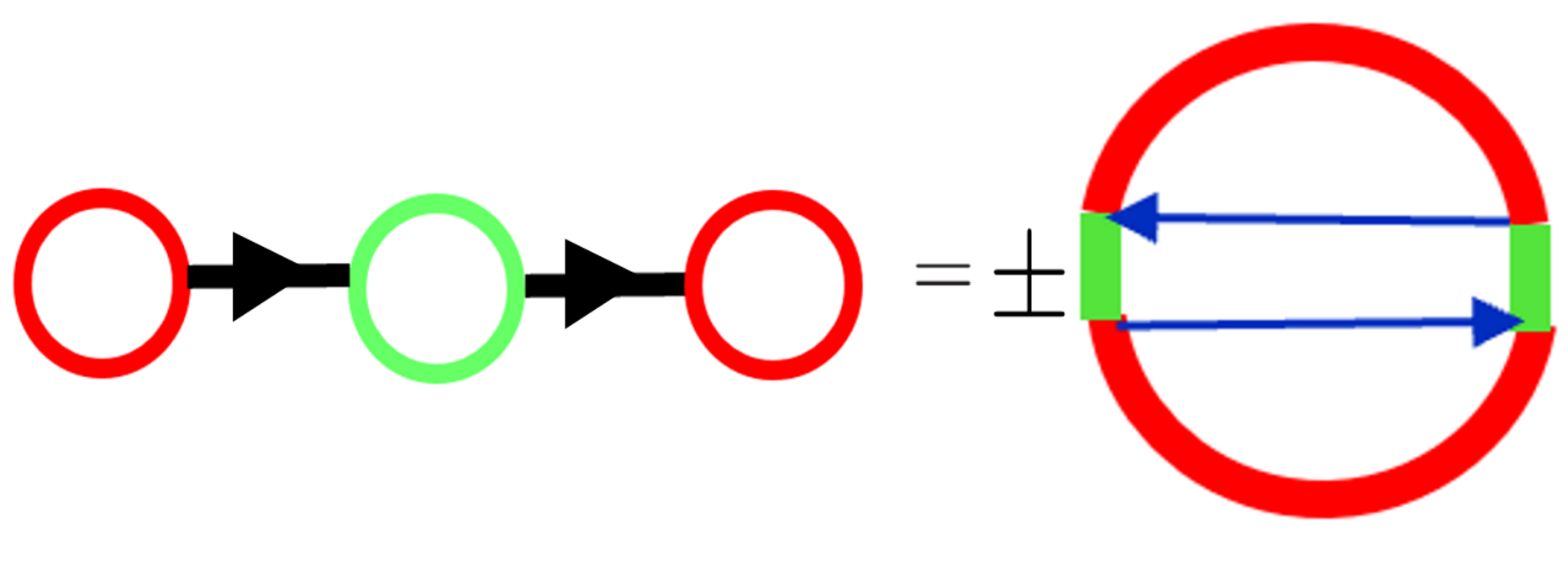}
\caption*{(d)}
\end{minipage}
\caption{Diagrams of tree operators }
\label{alltree}
\end{figure}

The tree operators made from $\mathcal{K}_{\textcolor{red}{2}}$ or $\mathcal{K}_{\textcolor{green}{2}}$ alone
and involved chains with both $\mathcal{K}_{\textcolor{red}{2}}$ and $\mathcal{K}_{\textcolor{green}{2}}$
are drawn in Figs.\ref{alltree} (a), (b) and (c), (d) respectively. All tree-operators are depicted as one
connected diagram which are called the connected operators.

\begin{figure}[H]
\centering
\begin{minipage}[c]{0.4\textwidth}
\centering
\includegraphics[height=2.6cm]{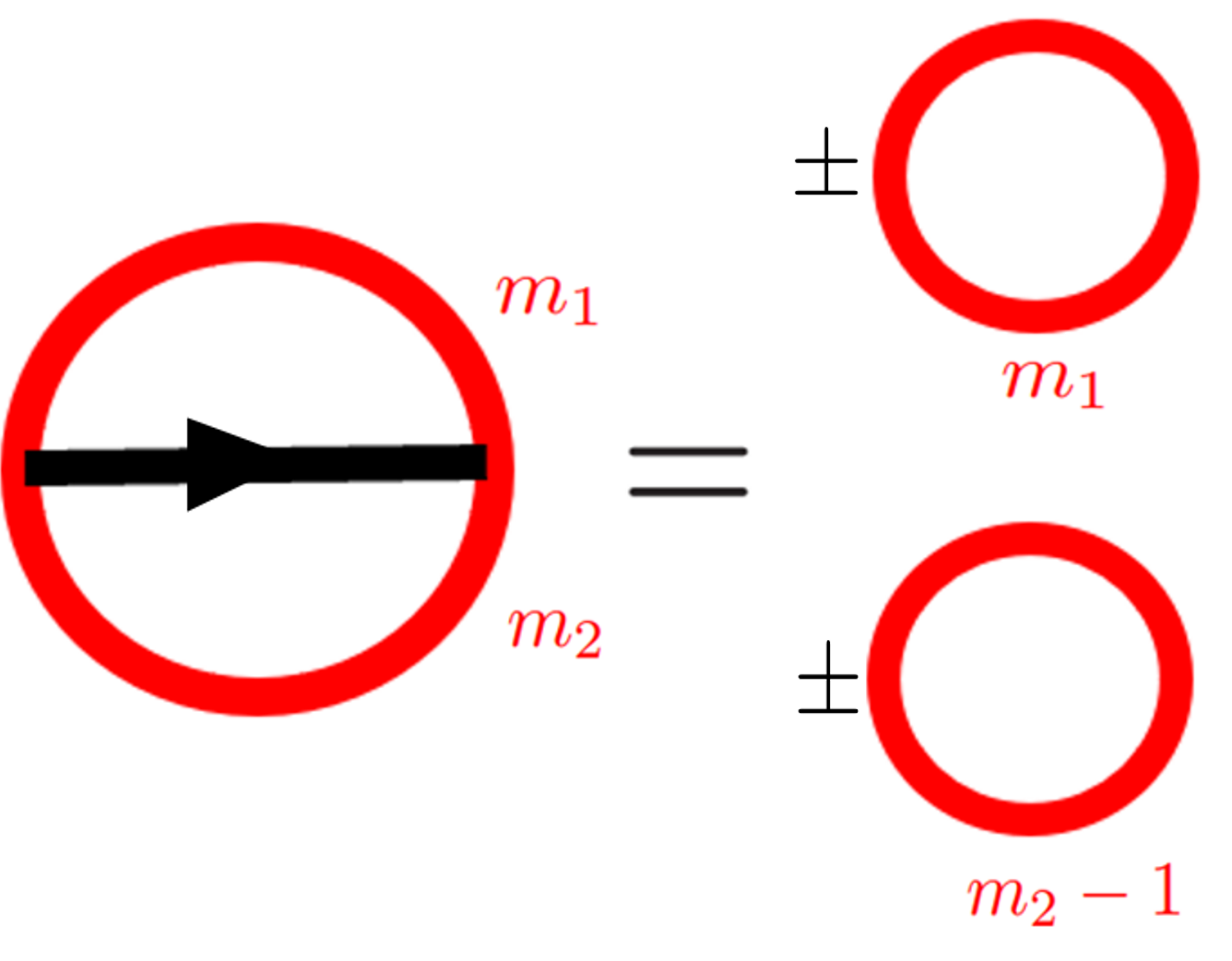}
\caption*{(a)}
\end{minipage}
\begin{minipage}[c]{0.4\textwidth}
\centering
\includegraphics[height=2.7cm]{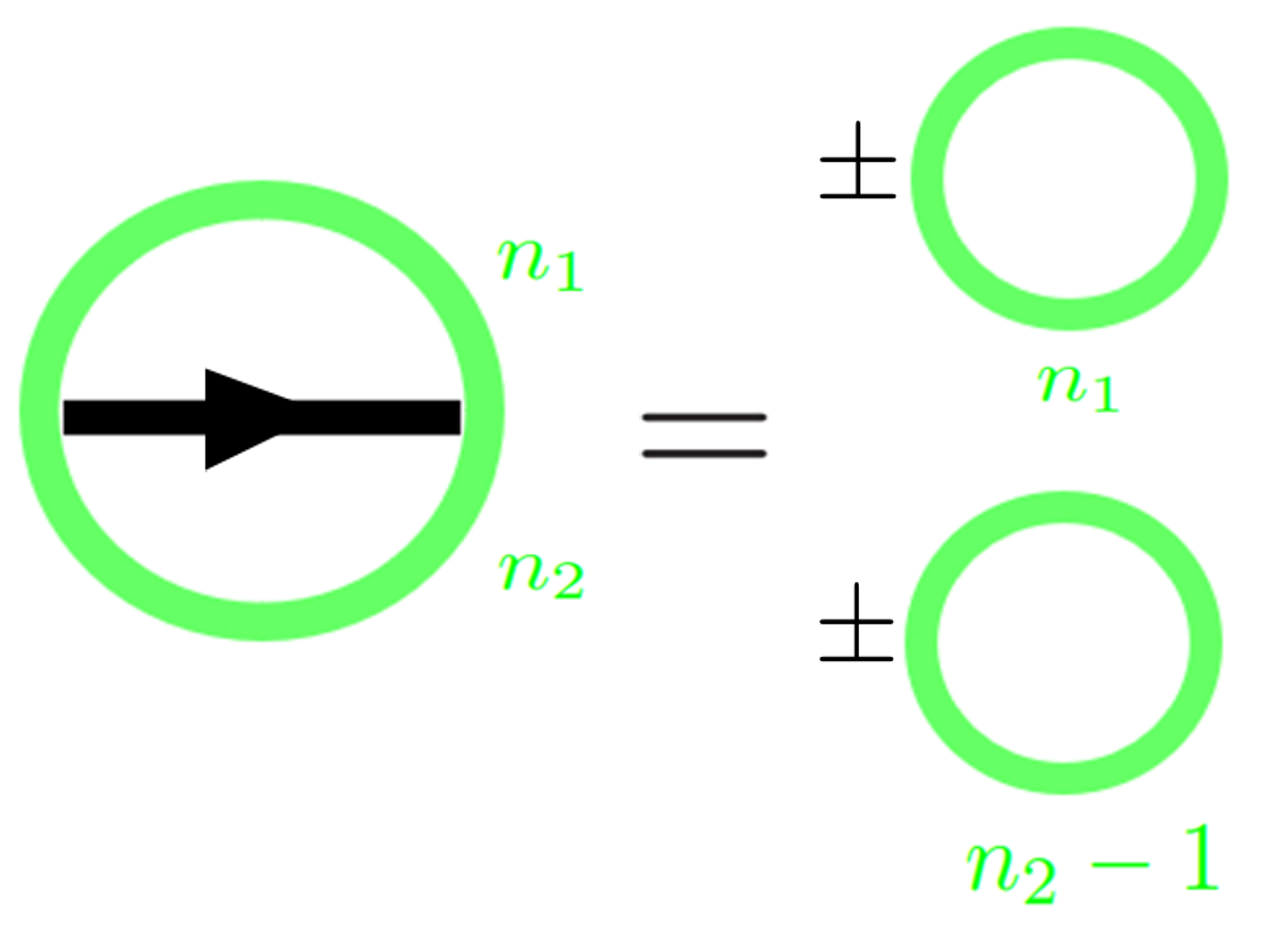}
\caption*{(b)}
\end{minipage}
\nonumber\\
\begin{minipage}[c]{0.4\textwidth}
\centering
\includegraphics[height=1.5cm]{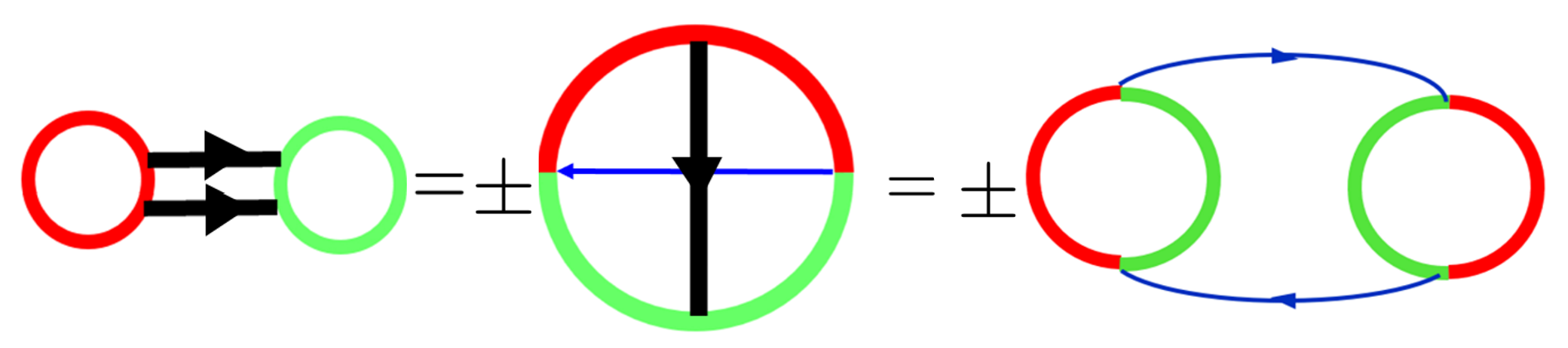}
\caption*{(c)}
\end{minipage}%
\begin{minipage}[c]{0.4\textwidth}
\centering
\includegraphics[height=2.15cm]{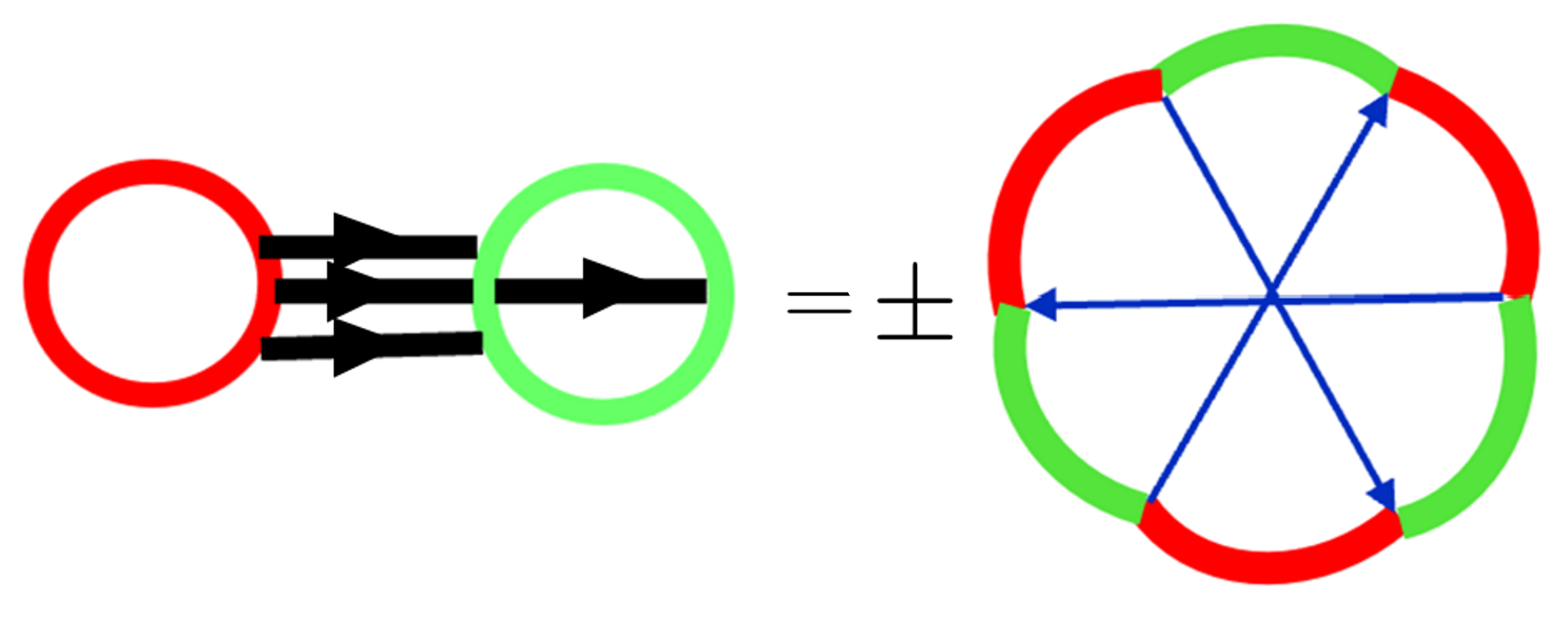}
\caption*{(d)}
\end{minipage}
\caption{Diagrams of loop operators }
\label{loopr=3}
\end{figure}

The loop operators made from $\mathcal{K}_{\textcolor{red}{2}}$ or $\mathcal{K}_{\textcolor{green}{2}}$ alone are drawn
in Figs.\ref{loopr=3} (a) and (b).  These operators which are depicted as disconnected collection of some diagrams are
called disconnected operators. When the loop operators involve both $\mathcal{K}_{\textcolor{red}{2}}$
and $\mathcal{K}_{\textcolor{green}{2}}$ are drawn in Figs.\ref{loopr=3} (c) and (d).

Similar to the case of the Aristotelian tensor model,
$\mathcal{K}_{\textcolor{red}{m}}=\mathcal{K}_{(12\cdots m)\otimes id\otimes id}$ and $\mathcal{K}_{\textcolor{green}{m}}=\mathcal{K}_{id\otimes(12\cdots m)\otimes id}$
are depicted as the red and green circles of length $m$, the thick color lines are
\begin{eqnarray*}
\includegraphics[height=2.75cm]{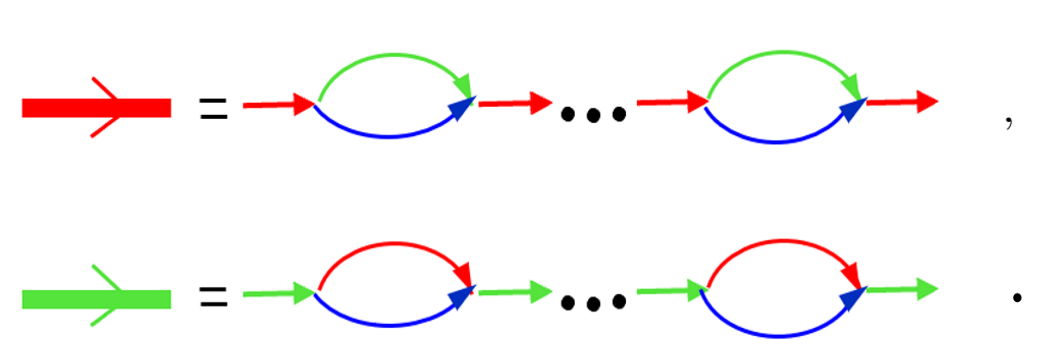}
\end{eqnarray*}

In terms of the keystones operators, connected tree and loop operators
in the ring, we introduce the fermionic Aristotelian tensor model
\begin{eqnarray}\label{aristotelian}
Z_{FA}
&=&\int d\Psi d\bar \Psi \exp( \Tr \bar \Psi \Psi+t_1^{(1)}\mathcal{K}_1^{(1)}+\sum_{{\textcolor{red}{k}}=2}^\infty {\textcolor{red}
{t^{(k)}_{k}}}\mathcal{K}_{{\textcolor{red}{k}}}^{\textcolor{red}{(k)}}
+\sum_{{\textcolor{green}{k}}=2}^\infty \textcolor{green}{t^{(k)}_{k}}\mathcal{K}^{\textcolor{green}{(k)}}_{\textcolor{green}{k}}
+\sum_{{\textcolor{blue}{k}}=2}^\infty \textcolor{blue}{t^{(k)}_{k}}\mathcal{K}^{\textcolor{blue}{(k)}}_{\textcolor{blue}{k}}
\nonumber\\&&
\quad\quad\quad\quad\quad+\sum_{{\textcolor{red}{k_1}},{\textcolor{green}{k_2}}=2}^\infty t_{{\textcolor{red}{k_1}},{\textcolor{green}
{k_2}}}^{(\textcolor{red}{k_1}+\textcolor{green}{k_2}-1)}\mathcal{K}_{{\textcolor{red}{k_1}},
{\textcolor{green}{k_2}}}^{(\textcolor{red}{k_1}+\textcolor{green}{k_2}-1)}
+\sum_{{\textcolor{red}{k_1}},{\textcolor{blue}{k_2}}=2}^\infty t_{{\textcolor{red}{k_1}},{\textcolor{blue}{k_2}}}^{(\textcolor{red}{k_1}
+\textcolor{blue}{k_2}-1)}\mathcal{K}_{{\textcolor{red}
{k_1}},{\textcolor{blue}{k_2}}}^{(\textcolor{red}{k_1}+\textcolor{blue}{k_2}-1)}
\nonumber\\&&
\quad\quad\quad\quad\quad+\sum_{{\textcolor{green}{k_1}},{\textcolor{blue}{k_2}}=2}^\infty t_{{\textcolor{green}{k_1}},{\textcolor{blue}{k_2}}}^{(\textcolor{green}{k_1}+\textcolor{blue}{k_2}-1)}\mathcal{K}_{{\textcolor{green}{k_1}},
{\textcolor{blue}{k_2}}}^{(\textcolor{green}{k_1}+\textcolor{blue}{k_2}-1)}
+\cdots)\nonumber\\
&=&\sum_{s=0}^{\infty}Z_{FA}^{(s)},
\end{eqnarray}
where the measure is induced by the norm~$\parallel \delta \Psi \delta \bar\Psi \parallel=
\delta \Psi^{\textcolor{red}{i}}_{\textcolor{green}{j_1},\textcolor{blue}{j_{2}}} \delta \bar{\Psi}_{\textcolor{red}{i}}^{\textcolor{green}{j_1},\textcolor{blue}{j_{2}}}$,
\begin{eqnarray}
Z_{FA}^{(s)}=\int d\Psi d\bar \Psi \exp(\Tr\bar \Psi\Psi)\cdot\sum_{l=0}^{\infty}\sum_{n_1+\cdots +n_l=s}
\frac{1}{l!}\langle\mathcal{K}_{\sigma_1}^{(n_1)}\mathcal{K}_{\sigma_2}^{(n_2)}\cdots\mathcal{K}_{\sigma_l}^{(n_l)}
\rangle t_{\sigma_1}^{(n_1)}t_{\sigma_2}^{(n_2)}\cdots t_{\sigma_l}^{(n_l)},
\end{eqnarray}
and the correlators~$\langle\mathcal{K}_{\sigma_1}^{(n_1)}\mathcal{K}_{\sigma_2}^{(n_2)}\cdots\mathcal{K}_{\sigma_l}^{(n_l)}
\rangle $ are defined by
\begin{eqnarray}
\langle\mathcal{K}_{\sigma_1}^{(n_1)}\mathcal{K}_{\sigma_2}^{(n_2)}\cdots\mathcal{K}_{\sigma_l}^{(n_l)}
\rangle =\frac{\int d\psi d\bar \psi \mathcal{K}_{\sigma_1}^{(n_1)}\mathcal{K}_{\sigma_2}^{(n_2)}
\cdots\mathcal{K}_{\sigma_l}^{(n_l)} \exp( \Tr\bar \Psi\Psi)}{\int d\psi d\bar \psi \exp( \Tr\bar \Psi\Psi)}.
\end{eqnarray}

Considering the deformation
$\delta \Psi=\dsum_{a=1}^{\infty}\dsum_{deg\alpha=a}t_{\alpha}^{(a)}\dfrac{\partial
\mathcal{K}_{\alpha}^{(a)}}{\partial \bar \Psi}$ of the integration variables in the integral (\ref{aristotelian}),
we may deduce the constraint for the partition function (\ref{aristotelian})
\begin{eqnarray}\label{dw}
(\tilde{D}-\tilde{W})Z_{FA}=0,
\end{eqnarray}
where
\begin{eqnarray}\label{operatorD}
\tilde{D}&=& \dsum_{a=1}^{\infty}\dsum_{\deg\alpha=a}at_{\alpha}^{(a)}\frac{\partial}{\partial t_{\alpha}^{(a)}},
\end{eqnarray}
\begin{eqnarray}\label{operator}
\tilde{W}&=&\dsum_{a,n=1}^{\infty}\dsum_{\deg\alpha=a}\sum_{\deg\sigma=n}\sum_{\deg\beta=n+a-1}\gamma_{\sigma,\alpha}^{\beta}
 t_{\alpha}^{(a)} t_{\sigma}^{(n)}\frac{\partial}{\partial t_{\beta}^{(n+a-1)}}+\mathcal{N}t_{id\otimes id\otimes id}^{(1)}
 %\textcolor{red}{N_1}\textcolor{green}{N_2}\textcolor{blue}{ N_3}
 \nonumber\\
&&+\dsum_{a=1}^{\infty}\dsum_{\deg\alpha=a}\sum_{k=1}^3\sum_{\substack{\beta_1,\cdots,\beta_k\\b_1+\cdots+b_k+1=a}}(1-\delta_{a,1})
\Delta_{\alpha}^{\beta_1,\cdots,\beta_k}t_{\alpha}^{(a)}
\frac{\partial}{\partial t_{\beta_1}^{(b_1)}}\cdots\frac{\partial}{\partial t_{\beta_k}^{(b_k)}},
\end{eqnarray}
and $\mathcal{N}=\textcolor{red}{N_{1}}\textcolor{green}{N_{2}} \textcolor{blue}{N_{3}}$,
the permutations $\alpha$, $\sigma$ and $\beta$ are taken from indices of connected operators in the ring.

The commutator of $\tilde{D}$ with $\tilde{W}$ is
\begin{eqnarray}\label{commu}
[\tilde{D},\tilde{W}]=\tilde{W}.
\end{eqnarray}

Straightforward calculation of the operators $\tilde D$ and $\tilde{W}$ acting on $Z_{FA}^{(s)}$
shows that
\begin{eqnarray}\label{dzs}
\tilde {D} Z_{FA}^{(s)}=sZ_{FA}^{(s)},
\end{eqnarray}
\begin{eqnarray}\label{increa}
\tilde{W}Z_{FA}^{(s)}=(s+1)Z_{FA}^{(s+1)}.
\end{eqnarray}
It indicates that the operators $\tilde D$ and $\tilde{W}$ are also the operators preserving and increasing
the grading, respectively. Thus the partition function can be realized by acting on elementary function
with exponents of the operator $\tilde{W}$
\begin{eqnarray}\label{exp}
Z_{FA}=\exp({\tilde{W}})\cdot 1.
\end{eqnarray}

We may also introduce the Virasoro constraints
\begin{eqnarray}\label{VconsFAM}
\tilde L_{m}Z_{FA}=0,
\end{eqnarray}
where the constraint operators $\tilde L_m$ are given by
\begin{eqnarray}
\tilde L_m=\tilde{W}^m(\tilde{W}-\tilde{D}), \ m\in \mathbb{N},
\end{eqnarray}
which yield the Witt algebra (\ref{witt}) and null 3-algebra (\ref{3Valg}).

By means of the Virasoro constraints (\ref{VconsFAM}), we can derive the compact expression of correlators
\begin{eqnarray}\label{corrf}
\left\langle \mathcal{K}_{\alpha_1}^{(a_1)}\cdots \mathcal{K}_{\alpha_{i}}^{(a_i)} \right\rangle=\frac{i!}{ m!\lambda_{(\alpha_1,\cdots,\alpha_i)}}\sum_{\tau}
P^{\tau(\alpha_1),\cdots,\tau(\alpha_i)}
\end{eqnarray}
where~$m=a_1+\cdots+a_i$, $\tau$ denotes all distinct permutations of $(\alpha_1,\cdots,\alpha_i)$ and~$\lambda_{(\alpha_1,\cdots,\alpha_i)}$
is the number of $\tau$ with respect to~$(\alpha_1,\cdots,\alpha_i)$,
$P^{\tau(\alpha_1),\cdots,\tau(\alpha_i)}$ are the coefficients of the
term
$t_{\tau(\alpha_1)}^{(a_1)}\cdots t_{\tau(\alpha_i)}^{(a_i)}$ in
$\tilde{W}^m=\sum_{i=1}^{m}\sum_{a_1+\cdots+a_i=m}\sum_{\deg\alpha_i=a_i}
P^{\alpha_1,\cdots,\alpha_i} t_{\alpha_1}^{(a_1)}\cdots t_{\alpha_{i}}^{(a_i)}+\cdots.$

Moreover, when particularized to the correlators $\langle (\mathcal{K}_{1} )^{i} \rangle$,
there are the exact results
\begin{eqnarray}
\langle (\mathcal{K}_{1} )^{i} \rangle=(\mathcal{N}-(i-1))\langle (\mathcal{K}_{1} )^{i-1} \rangle=\prod_{j=0}^{i-1}(\mathcal{N}-j).
\end{eqnarray}

Let us list some correlators (\ref{corrf}) by calculating $\tilde{W}^i$, $i=1,2,3$, respectively,
and represent them graphically with the help of the fermionic Wick theorem \cite{Zabrodin2013}.

(\rmnum{1})
\begin{eqnarray}\label{k1oper4}
%\left\langle \mathcal{K}_{1}\right\rangle =\left\langle \mathcal{K}^{(1)}_{id\otimes id\otimes id}\right\rangle=\mathcal{N}_3
%=\bar\Psi_{\textcolor{red}{i}}^{\textcolor{green}{j_1},\textcolor{blue}{j_2}}
%\Psi^{\textcolor{red}{i}}_{\textcolor{green}{j_1},\textcolor{blue}{j_2}}
\includegraphics[height=1.4cm]{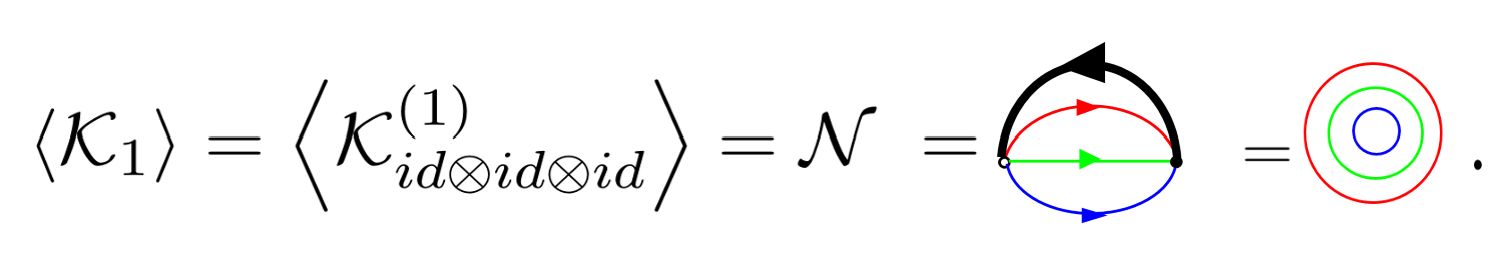}
\end{eqnarray}

(\rmnum{2})
\begin{eqnarray*}\label{k2oper4}
 &&\left\langle \mathcal{K}_{1}\mathcal{K}_{1}\right\rangle=\mathcal{N}^2-\mathcal{N}\nonumber\\
%&&\includegraphics[height=2cm]{K1k1-1}\includegraphics[height=2cm]{K1k1-2}
&&\includegraphics[height=2.25cm]{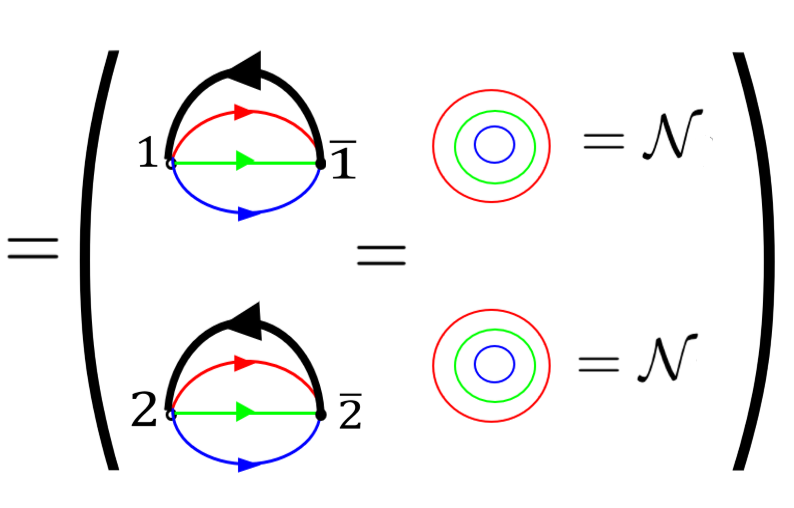}\includegraphics[height=2.2cm]{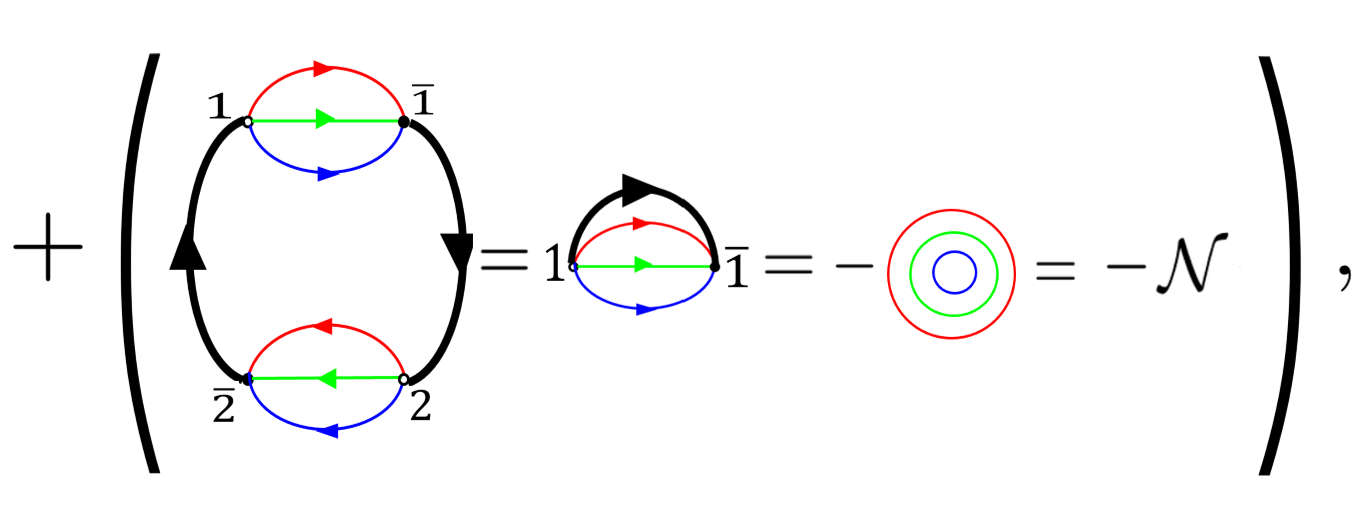}
\nonumber\\
  && \left\langle\mathcal{K}_{ \textcolor{red}{2}}\right\rangle=
   \left\langle \mathcal{K}_{id \otimes (12)\otimes (12) }^{(2)}\right\rangle
   =\textcolor{red}{N_1}\mathcal{N} -\textcolor{green}{N_2}\textcolor{blue}{N_3}\mathcal{N}
   \nonumber\\
&& \includegraphics[height=2.25cm]{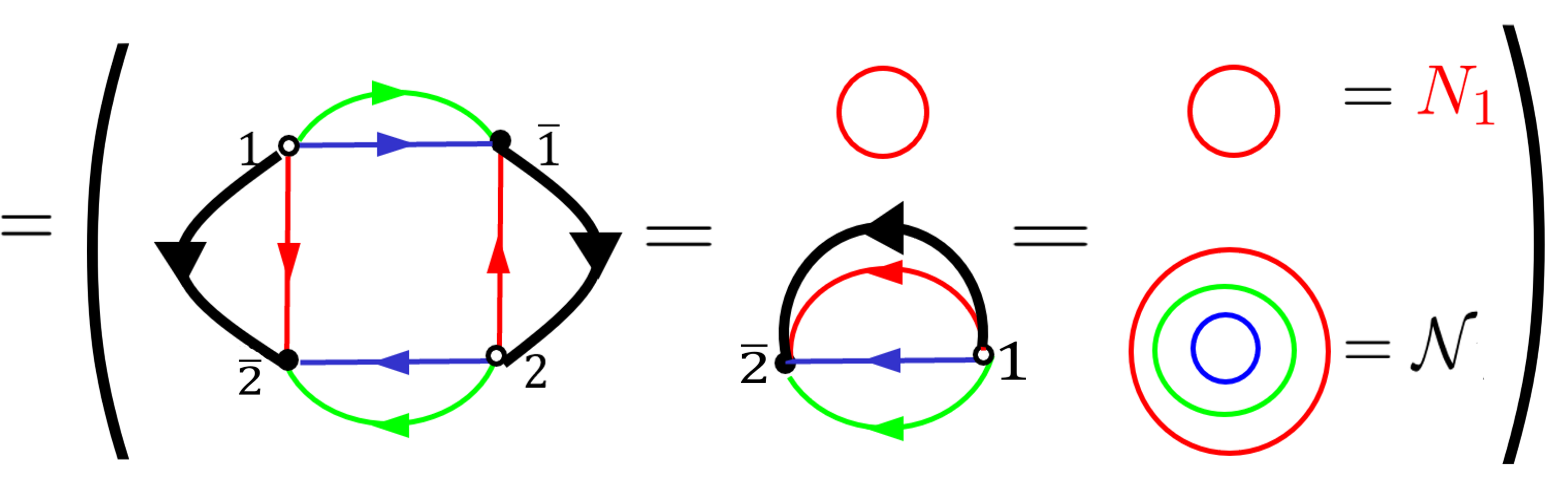}\includegraphics[height=2.2cm]{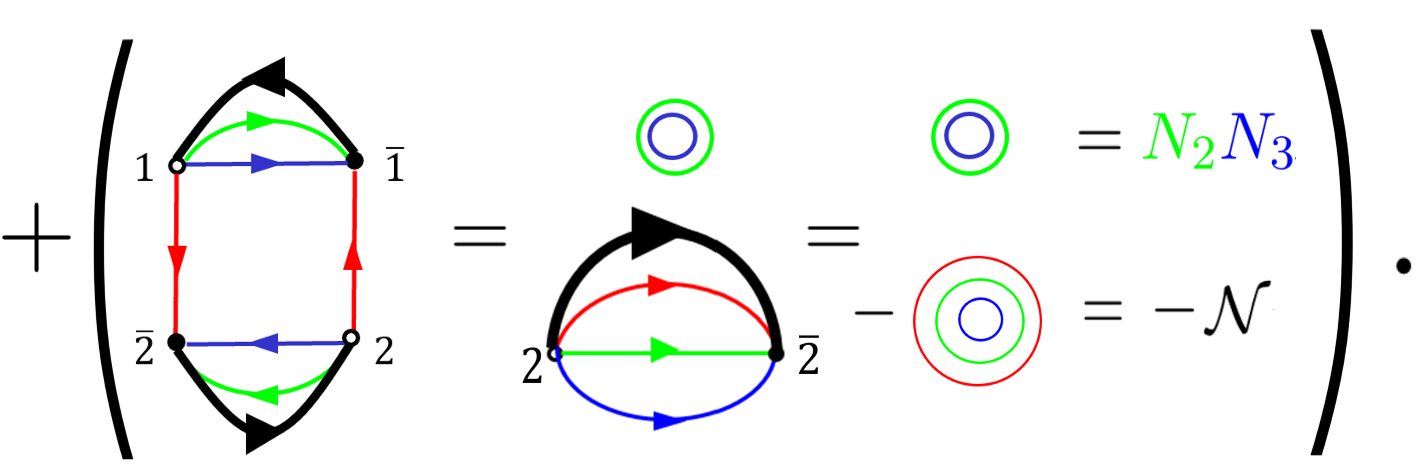}
\end{eqnarray*}

(\rmnum{3})
\begin{eqnarray*}
  &&\left\langle\mathcal{K}_{\textcolor{red}{3}}\right\rangle=\left\langle \mathcal{K}_{id \otimes (123)\otimes(123)}^{(3)}\right\rangle
  \nonumber\\
  && =
  \textcolor{red}{N_1}\cdot \textcolor{red}{N_1}\mathcal{N}-2\textcolor{red}{N_1}\cdot \textcolor{green}{N_2} \textcolor{blue}{N_3}\mathcal{N}
   +\textcolor{green}{N_2} \textcolor{blue}{N_3}\cdot\textcolor{green}{N_2} \textcolor{blue}{N_3}\mathcal{N}
   -\mathcal{N}^2+\mathcal{N}
   \nonumber\\
   &&\includegraphics[height=2.35cm]{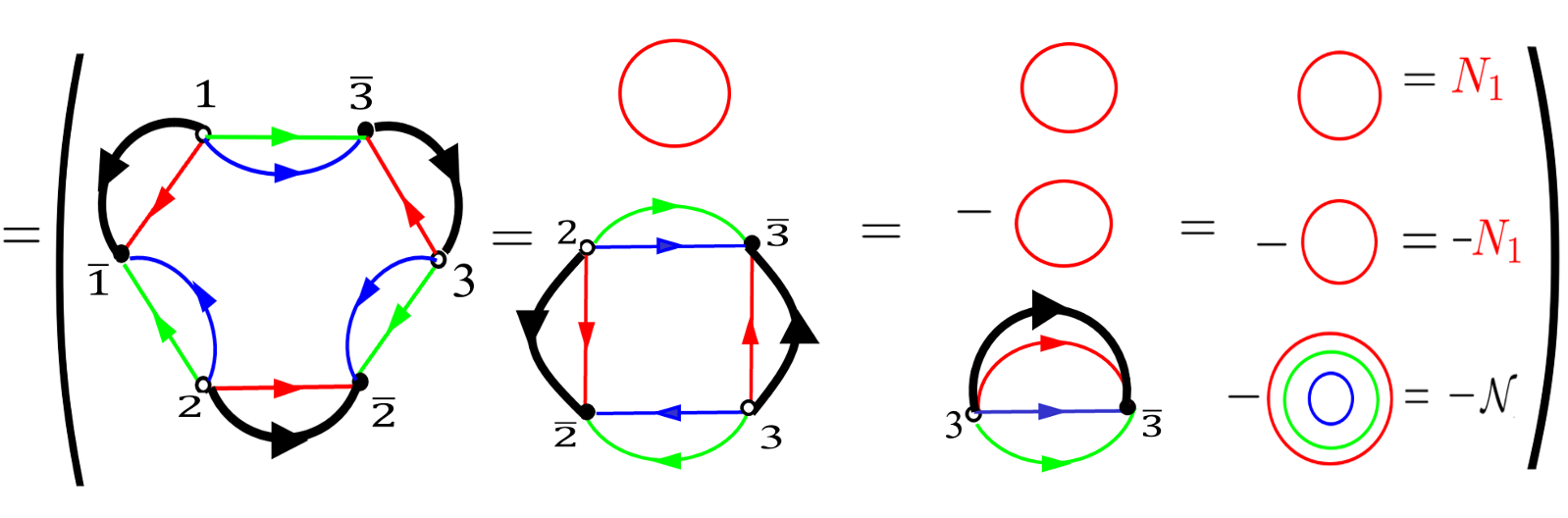}\includegraphics[height=2.4cm]{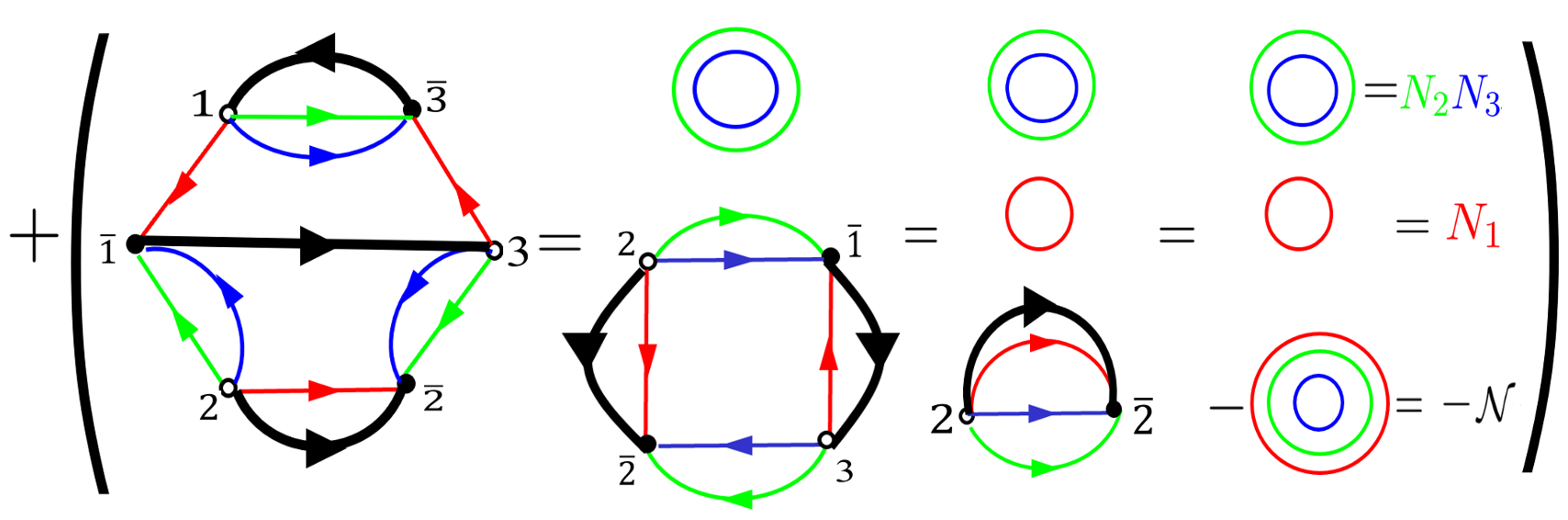}
   \nonumber\\
   &&\includegraphics[height=2.25cm]{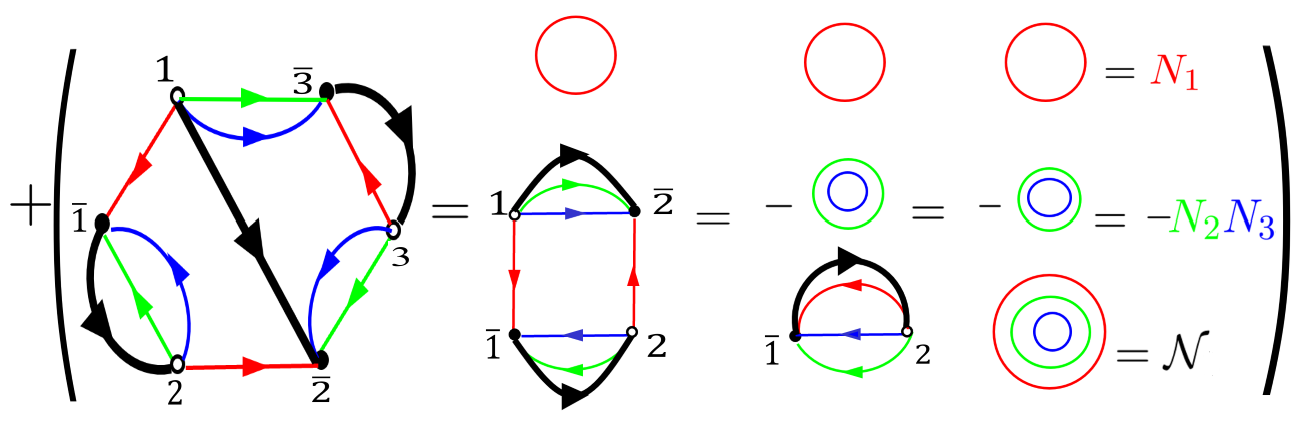}\includegraphics[height=2.25cm]{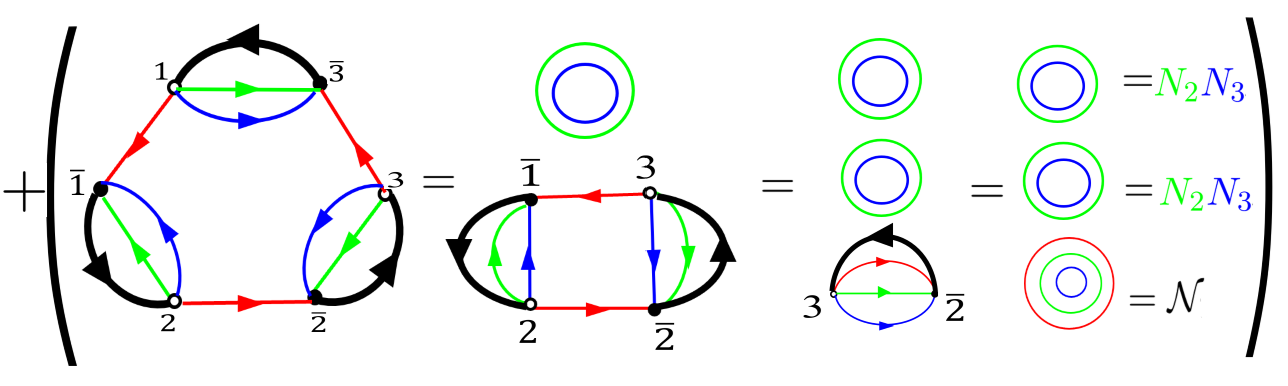}
    \nonumber\\
   &&\includegraphics[height=2.25cm]{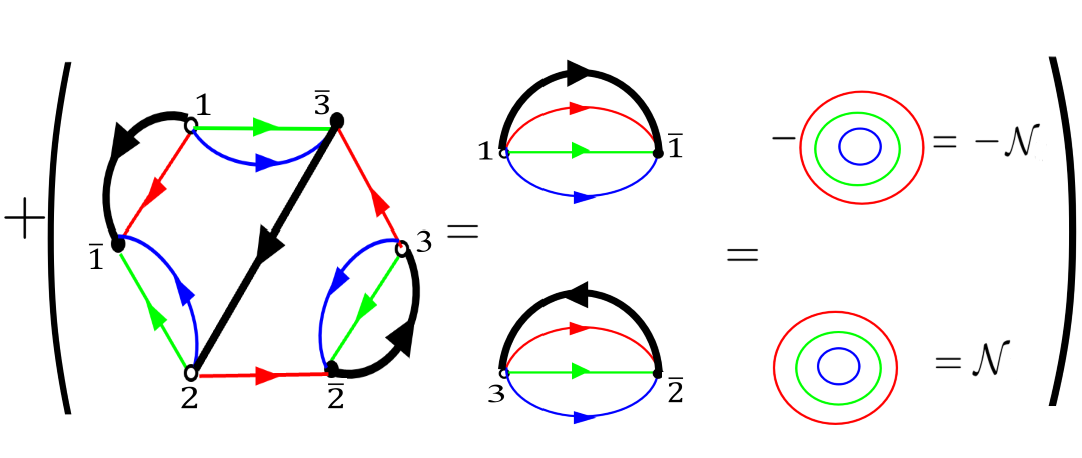}\includegraphics[height=2.25cm]{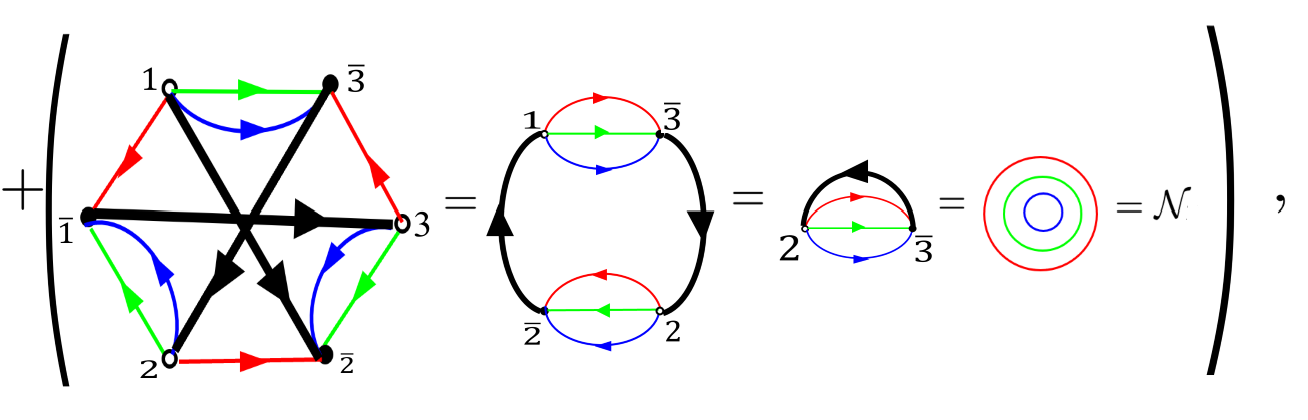}
\nonumber\\
&&
\left\langle \mathcal{K}_{id\otimes (12)\otimes (12)}^{(2)}\mathcal{K}_{id\otimes id\otimes id}^{(1)}\right\rangle
=\mathcal{N}(2\textcolor{red}{N_1}-2{\textcolor{green}{N_2}}{\textcolor{blue}{N_3}}
-\textcolor{red}{N_1}\mathcal{N}+{\textcolor{green}{N_2}}{\textcolor{blue}{N_3}}
\mathcal{N})\nonumber\\
&&
\includegraphics[height=2.85cm]{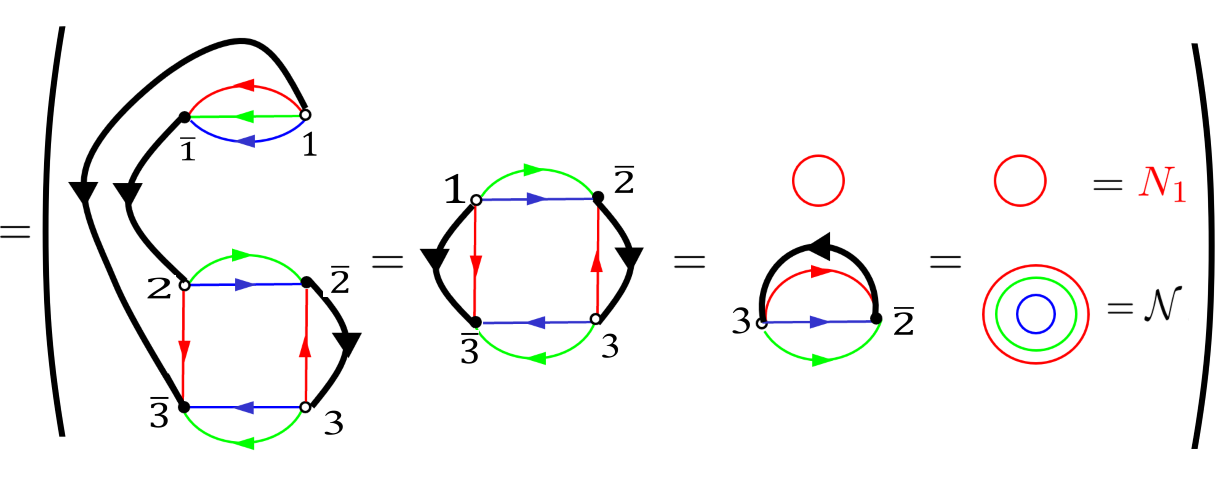}\includegraphics[height=2.7cm]{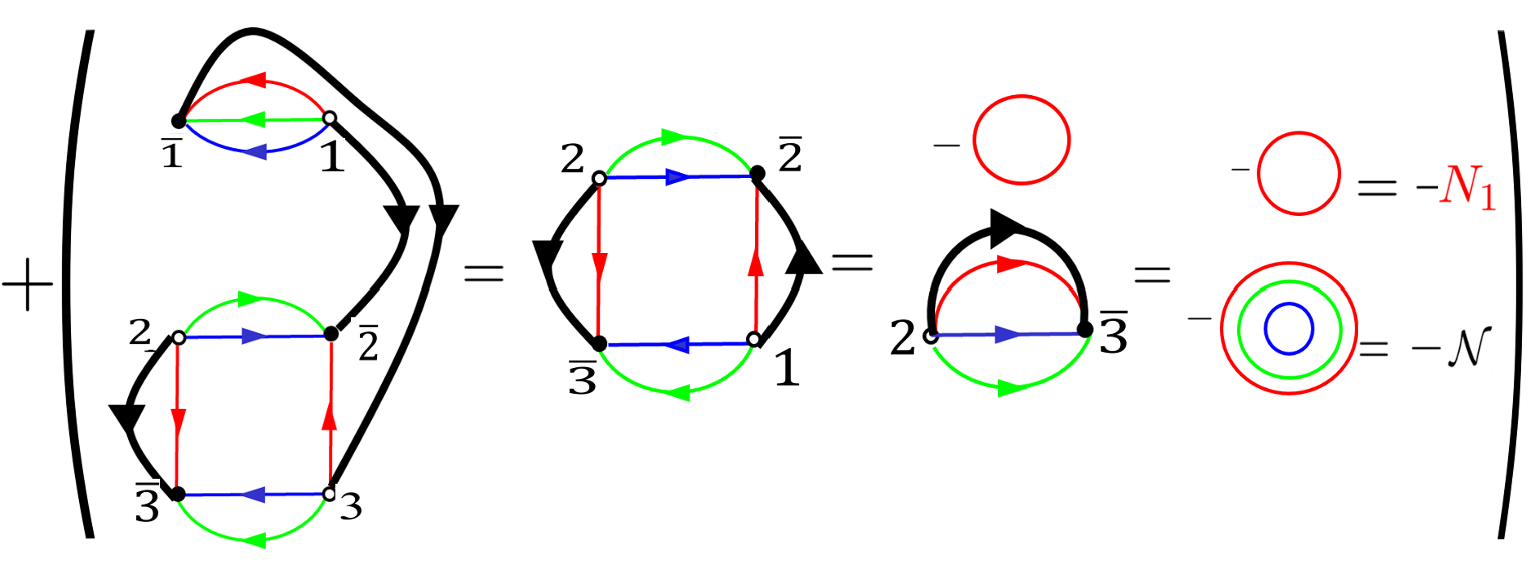}\nonumber\\
&&
\includegraphics[height=2.65cm]{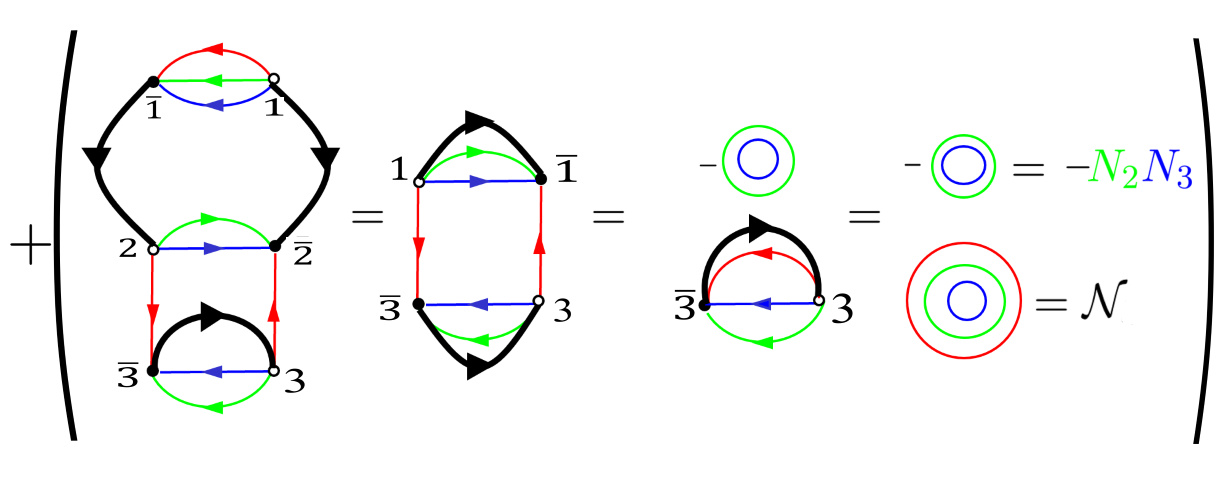}\includegraphics[height=2.55cm]{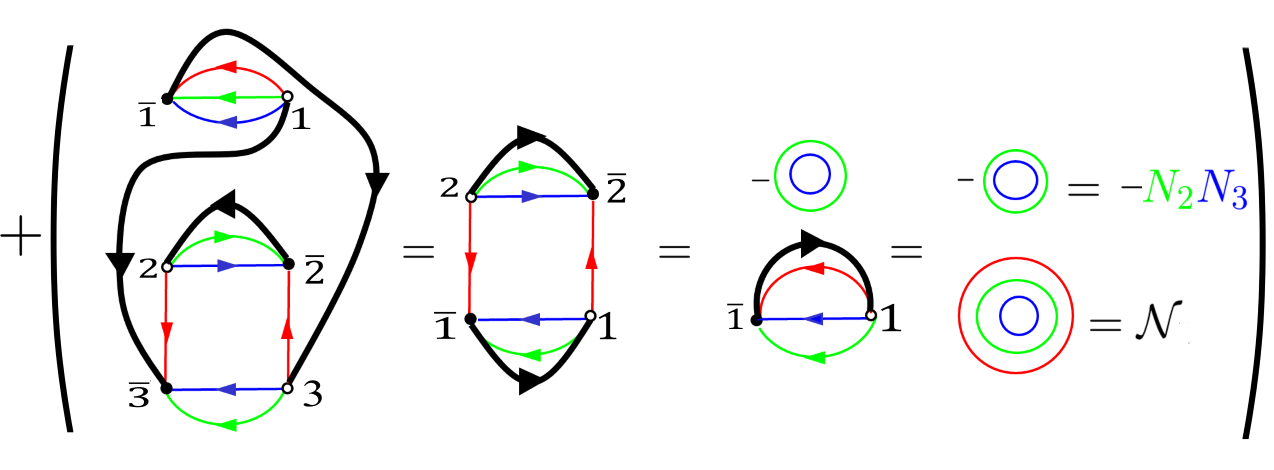}\nonumber\\
&&
\includegraphics[height=3.5cm,width=6cm]{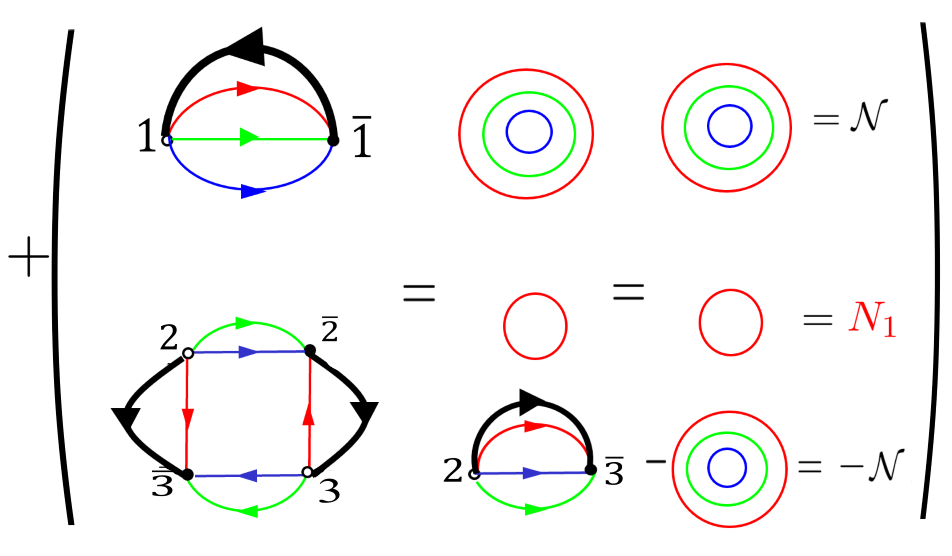}\includegraphics[height=3.5cm,width=6cm]{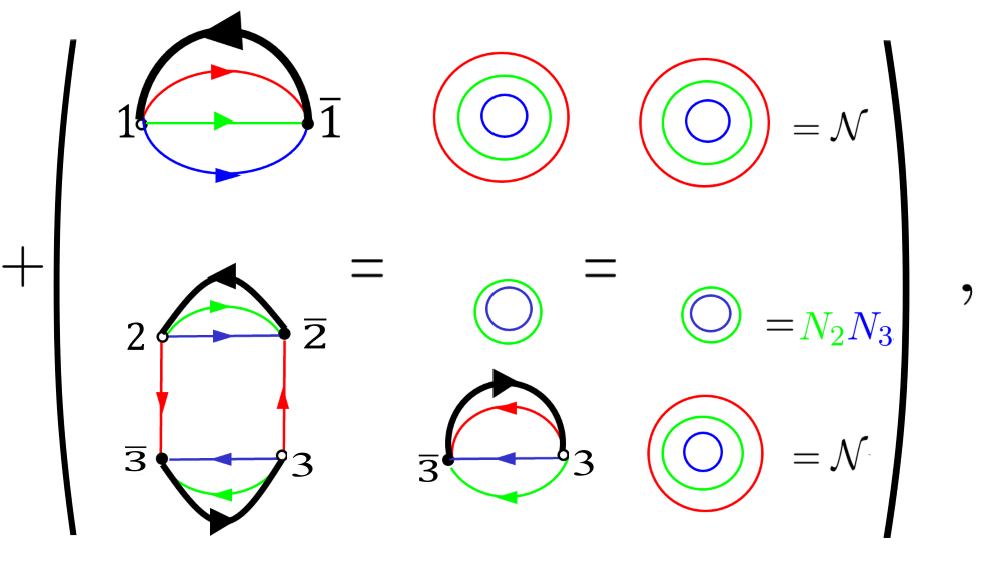}
\nonumber\\
&&\langle \mathcal{K}_1 \mathcal{K}_1 \mathcal{K}_1\rangle
=\mathcal{N}^3-3\mathcal{N}^2+2\mathcal{N}\nonumber\\
&&
\includegraphics[height=3.2cm,width=4.5cm]{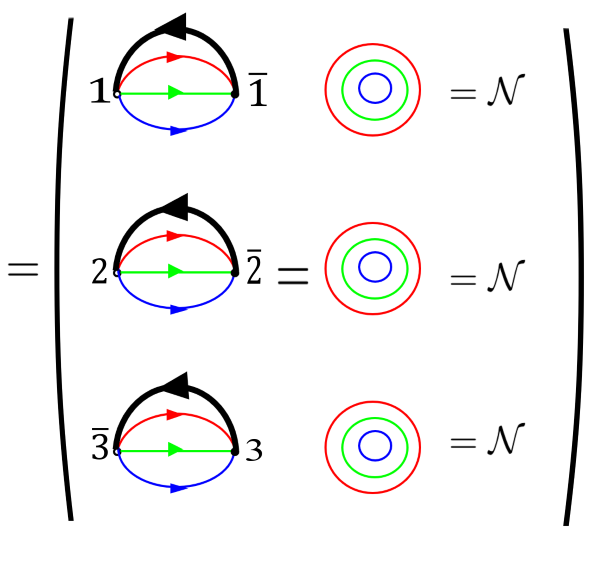}\includegraphics[height=3cm]{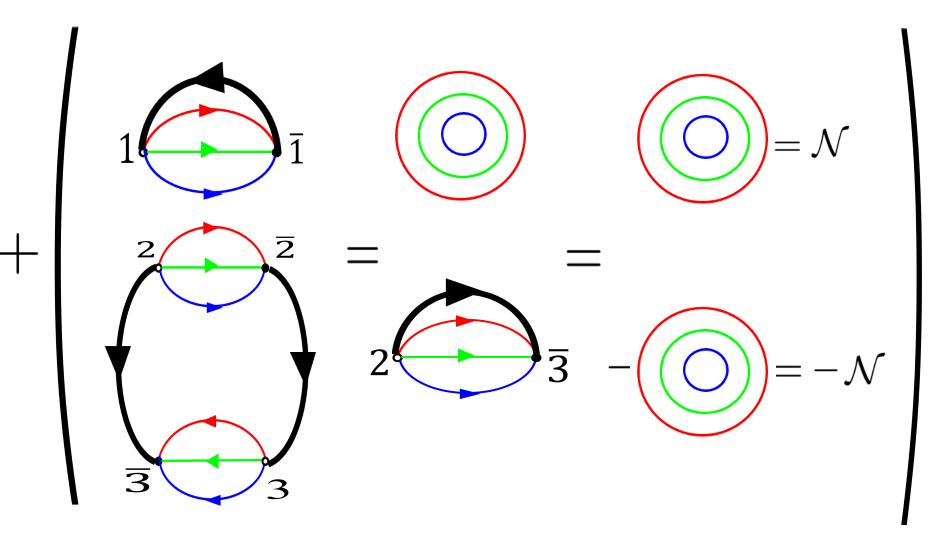}\nonumber\\
&&
\includegraphics[height=2.6cm]{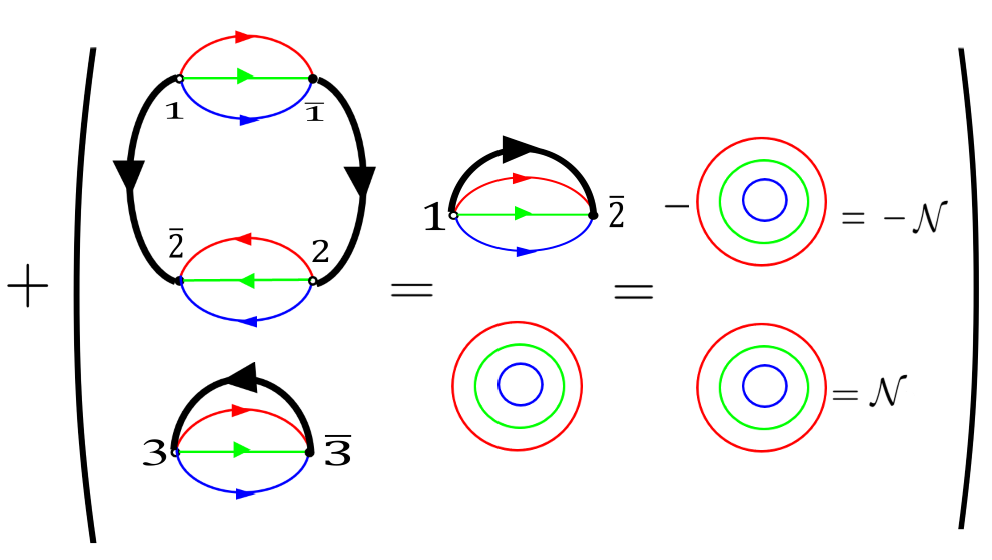}\includegraphics[height=2.6cm]{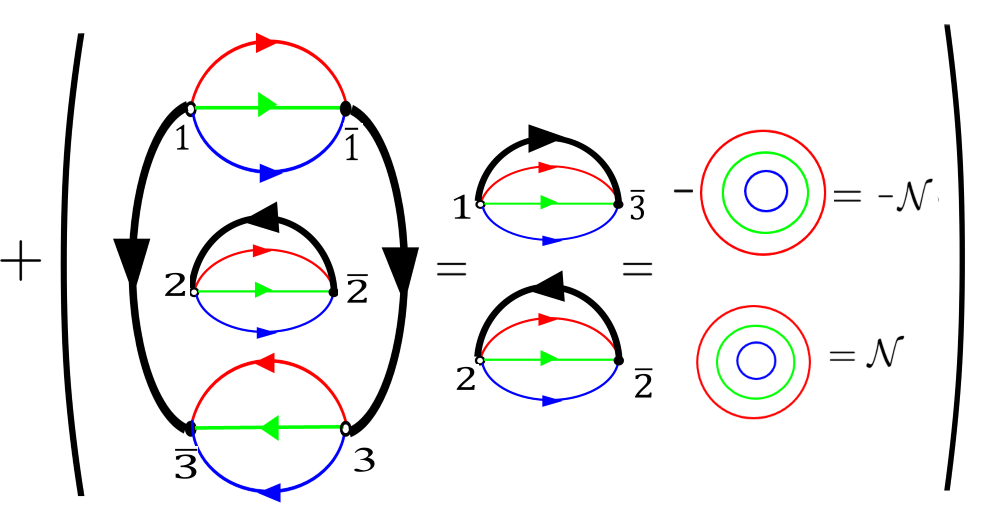}\nonumber\\
&&
\includegraphics[height=2.6cm]{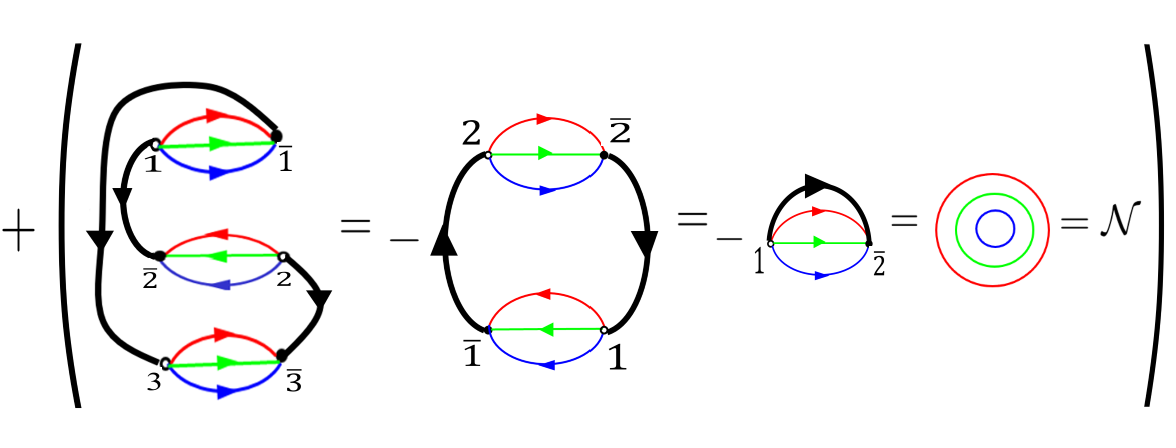}\includegraphics[height=2.6cm]{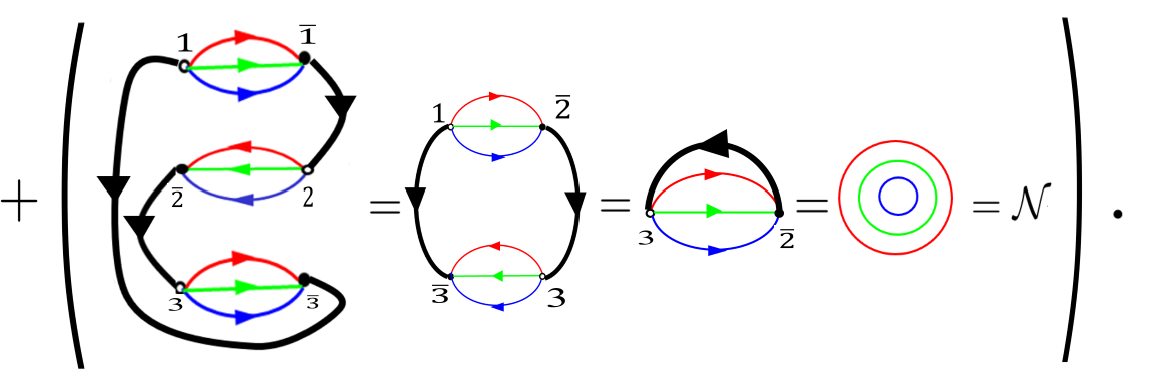}
\end{eqnarray*}

\section{$W$-representation and character expansion of the fermionic red tensor model}

The red tensor model was constructed in \cite{ItoyamaJHEP2017}.  It is indeed equivalent to a rectangular matrix model.
In order to generalize the red tensor model to the fermionic case, let us
consider the rank~$r$ fermionic tensors and introduce the gauge-invariant operators
\begin{eqnarray}\label{gauge-invariant}
\textcolor{red}{\mathcal{K}_n}&\equiv&\mathcal{K}_\sigma^{(n)}
\nonumber\\
&=&\bar\Psi_{\textcolor{red}{i}^{(1)}}^{{\textcolor{green}{j_1}^{(1)}},{\textcolor{blue}{j_2}^{(1)}},\cdots,{j_{r-1}}^{(1)}}
\Psi_{{\textcolor{green}{j_1}^{(1)}},{\textcolor{blue}{j_2}^{(1)}},\cdots,{j_{r-1}}^{(1)}}^{\textcolor{red}{i}^{(2)}}
\bar\Psi_{\textcolor{red}{i}^{(2)}}^{{\textcolor{green}{j_1}^{(2)}},{\textcolor{blue}{j_2}^{(2)}},\cdots,{j_{r-1}}^{(2)}}
\Psi_{{\textcolor{green}{j_1}^{(2)}},{\textcolor{blue}{j_2}^{(2)}},\cdots,{j_{r-1}}^{(2)}}^{\textcolor{red}{i}^{(3)}}
\nonumber\\
&&
\cdots
\bar\Psi_{\textcolor{red}{i}^{(n-1)}}^{{\textcolor{green}{j_1}^{(n-1)}},{\textcolor{blue}{j_2}^{(n-1)}},\cdots,{j_{r-1}}^{(n-1)}}
\Psi_{{\textcolor{green}{j_1}^{(n-1)}},{\textcolor{blue}{j_2}^{(n-1)}},\cdots,{j_{r-1}}^{(n-1)}}^{\textcolor{red}{i}^{(n)}}
\nonumber\\
&&
\cdot \bar\Psi_{\textcolor{red}{i}^{(n)}}^{{\textcolor{green}{j_1}^{(n)}},{\textcolor{blue}{j_2}^{(n)}},\cdots,{j_{r-1}}^{(n)}}
\Psi_{{\textcolor{green}{j_1}^{(1)}},{\textcolor{blue}{j_2}^{(1)}},\cdots,{j_{r-1}}^{(1)}}^{\textcolor{red}{i}^{(1)}},
\end{eqnarray}
where the gauge symmetry is $U(N_1)\otimes\cdots\otimes U(N_r)$, we take
$\sigma=(12\cdots n) \otimes id\otimes  \cdots \otimes id$ to be the simplest element of the double coset $\mathcal{S}_n^r=S_n\setminus S_n^{\otimes r}/S_n$.
It is clear that $\textcolor{red}{\mathcal{K}_n}$ are the connected operators.

We introduce the fermionic red tensor model
\begin{eqnarray}\label{red}
\textcolor{red}{Z}=\frac{\int d\Psi d\bar \Psi \exp[\mathcal{N}_r\Tr \bar\Psi\Psi +\sum_{{\textcolor{red}{n}}=1}^\infty
\frac{{\textcolor{red}{p_n}}}{\textcolor{red}{n}}{\textcolor{red}{\mathcal{K}_n}}]}{\int d\Psi d\bar\Psi\exp(\mathcal{N}_r\Tr\bar\Psi\Psi)},
\end{eqnarray}
where $\mathcal{N}_r=\textcolor{red}{N_{1}}\textcolor{green}{N_{2}}\textcolor{blue}{N_{3}}\cdots N_{r}$.

Following the same procedure as in the previous section, we may obtain
the corresponding operators preserving and increasing the grading
\begin{eqnarray}\label{redw2}
\textcolor{red}{\bar{D}}=\dsum_{\textcolor{red}{a}=1}^{\infty}\textcolor{red}{ap_a}\frac{\partial}{\partial \textcolor{red}{p_a}},
\end{eqnarray}
and
\begin{eqnarray}\label{redw1}
\textcolor{red}{\bar{W}}&=&
(\frac{\textcolor{red}{N_1}}{\mathcal{N}_r} -\frac{1}{\textcolor{red}{N_1}}) \sum_{{\textcolor{red}{b}}=1}^{\infty}{\textcolor{red}{b}}
{\textcolor{red}{p_{b+1}}}\dfrac{\partial }{\partial {\textcolor{red}{p_b}}}
+{\textcolor{red}{p_1}}
-\frac{1}{\mathcal{N}_r}\sum_{{\textcolor{red}{b_1,b_2}}=1}^{\infty}{\textcolor{red}{b_1}}
{\textcolor{red}{b_2}}{\textcolor{red}{p_{b_1+b_2+1}}}
\frac{\partial^2}{\partial {\textcolor{red}{p_{b_1}}}\partial {\textcolor{red}{p_{b_2}}}}
\nonumber\\
&&
-\frac{1}{\mathcal{N}_r}\sum_{{\textcolor{red}{b_1,b_2}}=1}^{\infty}{\textcolor{red}{(b_1+b_2-1)p_{b_1}p_{b_2}}}
\frac{\partial}{\partial
\textcolor{red}{p_{b_1+b_2-1}}}.
\end{eqnarray}

In deriving the above operators, we have considered the deformation
$\delta \Psi=\dsum_{a=1}^{\infty}\frac{\textcolor{red}{p_{a}}}{\mathcal{N}_r}
\frac{\partial \textcolor{red}{\mathcal{K}_{a}}}{\partial \bar \Psi}$ of the integration
variables in (\ref{red}), and used the similar cut and join operations with (\ref{cut}) and (\ref{join}).

Similarly, the partition function (\ref{red}) can be realized by the $W$-representation
\begin{eqnarray}\label{redwrep}
\textcolor{red}{Z}=\exp({\textcolor{red}{\bar{W}}})\cdot 1.
\end{eqnarray}

Moreover, there are the Virasoro constraints
\begin{eqnarray}\label{redvirasorom}
\bar L_m\textcolor{red}{Z}=0,
\end{eqnarray}
where the Virasoro constraint operators are $\bar L_m=\textcolor{red}{\bar{W}}^m(\textcolor{red}{\bar{W}}-\textcolor{red}{\bar{D}})
 $ which yield the Witt algebra (\ref{witt}) and null 3-algebra (\ref{3Valg}) as well.

Following the same procedure as in the fermionic matrix model,
we may derive the character expansion of (\ref{red}) from the Virasoro constraints (\ref{redvirasorom})
\begin{eqnarray}\label{redcharcter}
\textcolor{red}{Z}=\sum_R C_R\chi_R,
\end{eqnarray}
where
\begin{eqnarray}\label{C_S}
C_R=(\frac{-1}{\mathcal{N}_r})^{|R|}\frac{D_R(-\textcolor{green}{N_2}\textcolor{blue}{N_3}\cdots N_r)D_R(\textcolor{red}{N_1})}{d_R}.
\end{eqnarray}
It also coincides with the $\tau$-function (\ref{tau}) of the KP hierarchy.

Moreover there is the compact expression of correlators
\begin{eqnarray}\label{trivialcorre}
\left\langle \textcolor{red}{\mathcal{K}_{\alpha_1}}^{(a_1)}\cdots \textcolor{red}{\mathcal{K}_{\alpha_{i}}}^{(a_i)} \right\rangle
=\frac{i!}{ m!\lambda_{(\alpha_1,\cdots,\alpha_i)}}\sum_{\tau}
P^{\tau(\alpha_1),\cdots,\tau(\alpha_i)}\mathcal{N}_r^{k-l},
\end{eqnarray}
where $P^{\tau(\alpha_1),\cdots,\tau(\alpha_i)}$ are the coefficients of the term
$t_{\tau(\alpha_1)}^{(a_1)}\cdots t_{\tau(\alpha_i)}^{(a_i)}$ in
$\textcolor{red}{\bar{W}}^m$.

The first several exact correlators are
\begin{eqnarray}
&&\langle\textcolor{red}{\mathcal{K}_1}\rangle=1,\nonumber\\
&&\langle\textcolor{red}{\mathcal{K}_1}\textcolor{red}{\mathcal{K}_1}\rangle
=1-\mathcal{N}_r^{-1},\nonumber\\
%&&\langle\textcolor{red}{\mathcal{K}_1}\textcolor{red}{\mathcal{K}_1}\textcolor{red}{\mathcal{K}_1}\rangle
%=1-\mathcal{N}_r^{-1},\nonumber\\
&&\langle\textcolor{red}{\mathcal{K}_2}\rangle
=\textcolor{red}{N_1}\mathcal{N}_r^{-1}-\textcolor{red}{N_1}^{-1},\nonumber\\
&&\langle\textcolor{red}{\mathcal{K}_1}\textcolor{red}{\mathcal{K}_2}\rangle
=3\textcolor{red}{N_1}\mathcal{N}_r^{-1}
-3\textcolor{red}{N_1}^{-1}
-2\textcolor{red}{N_1}\mathcal{N}_r^{-2}
+2\textcolor{red}{N_1}^{-1}\mathcal{N}_r^{-1},\nonumber\\
&&\langle\textcolor{red}{\mathcal{K}_3}\rangle
=\textcolor{red}{N_1}^2\mathcal{N}_r^{-2}
+\textcolor{red}{N_1}^{-2}
+\mathcal{N}_r^{-2}
-3\mathcal{N}_r^{-1}.
\end{eqnarray}

%%%%%%%%%%%%%%%%%%%%%%%%%%%%%%%%%%%%%%%%%%%%%%%%%%%%%%%%%%%
\section{Conclusions}
%%%%%%%%%%%%%%%%%%%%%%%%%%%%%%%%%%%%%%%%%%%%%%%%%%%%%%%%%%%

We have analyzed the fermionic matrix model with complex Grassmann-valued~$N\times N$ matrices.
Similar to the Gaussian hermitian model case \cite{Shakirov2009},
there exist the operators $\hat D$ and $\hat W$ which preserve and increase the grading, respectively.
Thus the partition function can be realized by the $W$-representation.
The compact expression of correlators has been presented. In terms of the operators preserving
and increasing the grading, we may construct the Virasoro constraints such that the constraint operators
obey the Witt algebra and null 3-algebra. The remarkable feature is that the character expansion of
the partition function can be easily derived from such Virasoro constraints. It is known that
the superintegrability looks like a property $<character>\sim character$ \cite{MironovJHEP082018}.
The result of character expansion of the fermionic matrix model
shows that it is a $\tau$-function of the KP hierarchy.

We have given the fermionic Aristotelian and red tensor models which can also be realized
by the $W$-representations. Since there exist the operators preserving and increasing the
grading in these models, similarly, we may construct the desired Virasoro constraint
operators which lead to the higher algebraic structures. The compact expressions of correlators
in these models can be derived from such Virasoro constraints.
For the fermionic Aristotelian tensor model, it should be noted that we have to add the
infinite set of time variables in the partition function. It leads to that the formula \cite{macdonal}
$\chi_{\lambda/\mu}\{x^{(1)},\cdots,x^{(n)}\}=\sum_{(\nu)} \Pi_{i=1}^n \chi_{\nu^{(i)}/\nu^{(i-1)} }\{x^{(i)}\}$
with the finite set of variables $ x^{(1)},\cdots,x^{(n)}$ can not be applied in this case.
Hence we can not give its character expansion in the current state.

As the case of fermionic matrix model, there is one set of time variables in the fermionic red tensor model.
Thus similar character expansion of the partition function can be derived from the Virasoro constraints
with higher algebraic structures, which is a $\tau$-function of the KP hierarchy as well.
Finally, it should be pointed out that the Virasoro constraints with higher algebraic structures are also applicable
to the character expansions of the Gaussian hermitian matrix, complex matrix and red tensor models.
For further research, it would be interesting to study the cases of $\beta$ and $q,t$-deformed models

\section *{Acknowledgment}

This work is supported by the National Natural Science Foundation
of China (Nos. 11875194 and 11871350).

%%%%%%%%%%%%%%%%%%%%%%%%%%%%%%%%%%%%%%%%%%%%%%%%%%%%%%%%%%%%%%%%%%%%%%%%%%%%%%%

\end{document}